\documentclass[12pt,article]{amsart}

\usepackage{setspace}

\usepackage[margin=1in]{geometry}

\usepackage{graphicx,amssymb}
\usepackage{float}
\usepackage{caption}
\usepackage{algorithm}
\usepackage{algorithmic}
\usepackage{amsthm,color}
\usepackage{diagbox}
\usepackage{multirow}
\usepackage{makecell}
\usepackage{subcaption}
\usepackage{soul}
\usepackage{yhmath}
\usepackage{mathdots}
\usepackage{MnSymbol}

\usepackage[utf8]{inputenc}
\usepackage{multirow}

\usepackage{amsmath}
\usepackage{amssymb}
\usepackage{tabularx}
\usepackage{bbm,url}
\usepackage{graphicx}
\usepackage{grffile}
\usepackage{float}
\usepackage{caption}
\usepackage{subcaption}
\usepackage{adjustbox}
\usepackage{verbatim}
\graphicspath{{./Figures/}}

\usepackage[normalem]{ulem}

\usepackage{setspace}

\usepackage[table,xcdraw]{xcolor}
\usepackage{booktabs}

\usepackage{xcolor}
\usepackage{cleveref}

\usepackage{enumitem}
\usepackage{cite}
	
\usepackage[euler]{textgreek}

\newtheorem{thm}{Theorem}[section]

\newtheorem{prop}[thm]{Proposition}

\newtheorem{rem}[thm]{Remark} 
\newtheorem{assumption}[thm]{Assumption}

\numberwithin{equation}{section}

\providecommand{\keywords}[1]
{
  \small	
  \textbf{\textit{Keywords}---} #1
}

\title[Blind Deconvolution] {Blind Deconvolution of Nonstationary Graph Signals   over Shift-Invariant  Channels}
\author{Ali Zare, Yao Shi and Qiyu Sun
}%

\thanks{Zare, Shi, and Sun (Corresponding author) are with Dr. Yvette Kanouff School of Data, Mathematical, and Statistical Sciences, University of Central Florida, Orlando, Florida 32816, USA;
Emails: ali.zr1983@gmail.com; Yao.Shi@ucf.edu;   qiyu.sun@ucf.edu.
}
 
\begin{document}


\begin{abstract}  
In this paper, we investigate the blind deconvolution of nonstationary graph signals from noisy observations transmitted through an unknown shift-invariant channel, under the assumption that the covariance structure of the source  signals is known to the observer. Based on a phase retrieval framework, we develop a method to estimate the unknown channel and subsequently recover the source signals via frequency-domain Wiener filtering. We derive theoretical error bounds for both the proposed channel estimation  and the resulting blind deconvolution. The effectiveness of the proposed approach is 
validated through numerical experiments on cycle graphs and a real-world temperature dataset collected in the Brest region of France.
\end{abstract}

\maketitle

\section{Introduction}
\label{sec:intro}

Graph signal processing (GSP) provides a powerful framework for representing, processing, analyzing, and visualizing data defined on networks and other irregular domains \cite{Shuman2013, Ortega2018, Stankovic2019, Dong2020, Ortega2022, Isufi2024}. In this paper, we consider the problem of reconstructing a collection of nonstationary graph signals
${\bf x}_m, 1\le m\le M$, up to a sign
ambiguity, from noisy observations \begin{equation}\label{equ:model} \mathbf{y}_m = \mathbf{H}\mathbf{x}_m + \mathbf{n}_m, \qquad 1 \leq m \leq M, \end{equation} where $\mathbf{H}$ is an unknown shift-invariant graph filter modeling the transmission channel, and $\mathbf{n}_m$, $1 \leq m \leq M$, denotes additive white Gaussian noise with unknown variance. The proposed  blind deconvolution scheme operates under the assumption that the covariance information of the source signals 
is available to the observer.

In GSP, a channel refers to the medium through which a graph signal propagates from a transmitter to a receiver on the network. In this paper, we focus on linear channels ${\bf H}$
 that are shift-invariant with respect to a graph shift, such as the graph Laplacian; see \eqref{shiftinvariantfilter.Fourier}.
  Channel estimation is a critical component in a wide range of communication systems, including wireless communication, radar, audio processing, and medical imaging, as it enables the receiver to compensate for channel-induced distortions and mitigate or eliminate undesired effects on the source signals after measurement \cite{li2002, colieri2002study-ofdm, kofidis2013, zheng2022, singh2025, liu2014}. 
A wide range of channel estimation techniques has been developed for various communication systems, broadly classified into 
frequency-domain and time-domain approaches \cite{li2002, colieri2002study-ofdm, kofidis2013, Ye2025, Ye2025b, zheng2022}, as well as more recent machine-learning-based methods
\cite{ 
soltani2019, hu2021,   comm_NeuNet_MIMO-OFDM, mashdadi2021, ye2024_GNN,  
yang2020}. 
Based on the spectral characterization  of shift-invariant channels in \eqref{shiftinvariantfilter.Fourier}, we adopt a spectral approach to estimate the shift-invariant channel ${\bf H}$ in this paper; see Sections \ref{channelestiamtion.section} and \ref{channelestiamtion2.section}.

 Phase retrieval  arises in a wide range of applications, including X-ray crystallography, astronomy, optical computing and diffraction imaging, and it has been extensively studied in random modeling frameworks
\cite{Fienup82, jaganathan2016, Fannjiang2020, chen2022, chen2020, cheng2019, dong2023}.
The blind deconvolution problem considered in this paper closely resembles the classical phase retrieval problem,
as negated  signals $-{\bf x}_m, 1 \le m \le M$, preserve the same covariance structure and produce identical noisy observations ${\bf y}_m$  when passed through a negated channel $-{\bf H}$.
To recover the frequency responses of the channel $\mathbf{H}$, we utilize a phase retrieval procedure proposed in \cite{Li2017}, which first estimates the magnitude of the frequency responses and subsequently infers the signs;
see Algorithm \ref{alg1:channel-est}.  This recovery is guaranteed under the conditions: (i) all frequency responses of the shift-invariant channel ${\bf H}$ are nonzero, and (ii) the source graph is connected; see Proposition \ref{frequencynormestimate.prop} and Theorem \ref{population.uniqueness.thm}.
 Here, the source graph is constructed from the spectral covariance matrix of  
the source signals, and its connectedness implies that the source signals must be nonstationary and do not have any stationary frequency components; see Assumption \ref{mainassumption}.

Based on the estimation of the shift-invariant channel ${\bf H}$ developed in Sections~\ref{channelestiamtion.section} and \ref{channelestiamtion2.section}, we propose a blind deconvolution scheme to recover the source signals; see Section~\ref{recovery.subsection}. The proposed blind deconvolution  approach is essentially the Wiener filtering procedure applied independently at each frequency, without exploiting correlations of the source signals across different frequencies; see \eqref{blindreconstruction.eq}. Furthermore, under the assumption that the shift-invariant channel preserves all frequency components, we prove that, with high probability, the signals recovered by our blind deconvolution  scheme are close, up to a global sign ambiguity, to those obtained by applying the Wiener filter independently at each frequency using the true channel statistics; see \eqref{blinddeconvolutionsingalestimate} and Theorem~\ref{tildex.thm}.

The remainder of this paper is organized as follows. Section~\ref{sec:GSP} provides a brief overview of the fundamental concepts of  GSP. Section~\ref{sec:G-Channel} presents a phase-retrieval-based approach to estimate shift-invariant graph channels and introduces a blind deconvolution algorithm to  recover nonstationary graph signals. Section~\ref{demonstration.section} evaluates the performance of the proposed method through numerical experiments on cycle graphs and a real-world temperature dataset collected in the Brest region of France.

\section{Preliminaries in graph signal processing}\label{sec:GSP}

In this section, we recall some  preliminaries in GSP 
\cite{Shuman2013, Ortega2018, 
Dong2020, Ortega2022, 
Isufi2024}.

\begin{enumerate} 

\item \textit{Underlying graph}:
    Graphs provide a flexible tool to model the underlying topology of networks and irregular domains. In this paper, we assume that the underlying graph ${\mathcal G}$
    that models the topological structure of our dataset  is undirected, unweighted, and finite. We denote its set of vertices by $V$, its set of edges by $E\subset V\times V$ and its order  by $N\ge 1$.

 \item \textit{Path and traversal}: 
 A {\em path} on the  graph ${\mathcal G}:=(V, E)$ is a finite sequence of edges which joins a sequence of distinct vertices, 
 and a {\em traversal} on the graph ${\mathcal G}$ visits all vertices in the graph
with a unique path existing between every pair of distinct vertices.

\item \textit{Graph adjacency and Laplacian matrices}: The adjacency matrix ${\bf A} = [A(i,j)]_{i,j\in V}$ of the  graph ${\mathcal G}:= (V, E)$ has its entries $A(i,j)$ equal to one
whenever  there is an edge to connect vertices $i$ and $j$, and zero otherwise. The
graph Laplacian ${\bf L}$ is defined as the difference between the 
degree matrix ${\bf D}$ and the adjacency matrix ${\bf A}$  (i.e.,
${\bf L} = {\bf D} - {\bf A}$), where  ${\bf D}$ is a diagonal matrix with diagonal entries
$d(i) = \sum_{j\in V} A(i,j)$ for $i \in V$. The adjacency matrix ${\bf A}$ of a graph
encodes its connectivity structure, while the Laplacian
${\bf L}$ is  linked to its numerous functional properties.

\item \textit{Graph shift}: A {\em graph shift} on the graph ${\mathcal G}:=(V, E)$,
represented by matrix $\mathbf{S} =  [S(i,j)]_{i,j \in V}$, has entries satisfying
$S(i,j) = 0$ unless $i=j$  or $(i, j)\in E$.
The concept of graph shift
plays a similar role in GSP as the one-order delay does in classical signal processing.  
Illustrative examples are
 the adjacency matrix ${\mathbf A}$, 
   the Laplacian matrix ${\mathbf L}$ 
  and their variants \cite{Ortega2022, Emirov2022}.
In this paper, we fix a graph shift ${\mathbf S}$ that is symmetric and real-valued, and has  {\em distinct} eigenvalues.
Under the above assumption, we can find an orthogonal matrix
$\mathbf{U} = [{\bf u}_1, \ldots, {\bf u}_N]$ and a diagonal matrix
${\pmb \Lambda} = \operatorname{diag}[\lambda(1), \ldots, \lambda(N)]$ such
that
\begin{equation}\label{eigendecomposition.def}
\mathbf{S}
= \mathbf{U}{\pmb \Lambda}\mathbf{U}^{\top}
= \sum_{n=1}^N \lambda(n)\, {\bf u}_n {\bf u}_n^{\top}.
\end{equation}
Moreover, the diagonal matrix ${\pmb \Lambda}$ contains the distinct eigenvalues $\lambda(1), \ldots, \lambda(N)$ of
$\mathbf{S}$ on its diagonal  ordered such that $|\lambda(1)| \le \cdots \le |\lambda(N)|$.

  \item \textit{Graph signal}: A graph signal $\mathbf{x}: V \rightarrow {\mathbb R}$
on graph ${\mathcal G}$ assigns some 
value  $ x(i)$ to vertex  $i\in V$, and is  represented by a vector ${\bf x}=[x(i)]_{i\in V}$ with
the vertex set $V$ 
as its index set. Denote the energy of a graph signal ${\bf x}$ by
$\|{\bf x}\|_2:=(\sum_{i\in V} |x(i)|^2)^{1/2}$.

\item \textit{Graph Fourier transform}: 
Based on the eigendecomposition
\eqref{eigendecomposition.def}  of the graph shift ${\bf S}$, we define the  {\em graph Fourier transform} (GFT) of a graph signal $\mathbf{x}$  by
\begin{equation}\label{GFT.def}
\widehat{\mathbf{x}}: = {\mathbf U}^{\top}\mathbf{x} =:[{\widehat x}(1), {\widehat x}(2), \ldots, {\widehat  x}(N)]^\top, \end{equation}
and set $\lambda(n)$ and ${\bf u}_n, 1\le n\le N$, as
its frequencies and modes of variation, respectively \cite{
Ortega2022, 
 Emirov2022, Chen2023}. 
By the orthogonality of ${\bf U}$, we have the following representation
of a graph signal ${\bf x}$ 
using various modes of variation,
$${\bf x}=\sum_{n=1}^N ({\bf u}_n^\top {\bf x}) {\bf u}_n=\sum_{n=1}^N {\widehat x}(n) {\bf u}_n.
$$
GFT is one of the fundamental tools in GSP
to decompose graph signals into different frequency
components and  represent graph signals with
regularity effectively using various modes of variation.

\item \textit{Graph channel}:
A  {\em graph channel} on graph ${\mathcal G}$ can be represented by a graph filter ${\bf H}$
 that linearly maps a graph signal ${\bf x}$ to another graph signal ${\bf H}{\bf x}$, and it is usually described by a matrix  ${\bf H}=[H(i,j)]_{i,j\in V}$.
 Graph filters and their implementations play  pivotal roles in GSP, and they have been used in denoising, smoothing, estimation  and many other engineering applications.

\item \textit{Shift-invariant graph channel}:
We say that a graph  channel $\mathbf{H}$ is  {\em shift-invariant} if
it commutes with the graph shift ${\bf S}$, i.e., 
$\mathbf{S}\mathbf{H} = \mathbf{H}\mathbf{S}$.
For a shift-invariant channel ${\bf H}$, we  have
\begin{equation}\label{shiftinvariantfilter.Fourier}  {\bf H} ={\bf U}{\pmb \Gamma}{\bf U}^\top
\end{equation}
for some diagonal matrix ${\pmb \Gamma}={\rm diag} [\gamma (1), \ldots, \gamma (N)]$,  with its diagonal entries considered as the  frequency responses of the shift-invariant channel $\mathbf{H}$ \cite{ Sandryhaila2013, Sandryhaila2014, Emirov2022}.
For the filtering procedure \eqref{equ:model}, the noisy observation
$\mathbf{y}$ is obtained from the source signal $\mathbf{x}$ passing through
the unknown shift‑invariant channel $\mathbf{H}$ with additive white noise ${\bf n}$,
\begin{equation}\label{xy.time}
\mathbf{y} = \mathbf{H}\mathbf{x} + \mathbf{n}.\end{equation}
Applying the GFT to both sides of the above expression yields
\begin{equation}\label{xy.fourier}
\widehat{\bf y}
= {\pmb \Gamma}\, \widehat{\bf x}
  + \widehat{\bf n},
\end{equation}
where  the diagonal matrix ${\pmb \Gamma}$ is given in
\eqref{shiftinvariantfilter.Fourier} that has frequency responses of
the channel ${\bf H}$ as its diagonal entries. 

\item \textit{Nonstationary graph signal}: Stationarity is a cornerstone of many signal analysis methods.
A random graph signal ${\bf x}$ is said to be {\em stationary} if it has zero mean ${\mathbb E}({\bf x})= \mathbf{0}$ and its covariance matrix
${\rm cov} ({\bf x})= {\mathbb E} ( ({\bf x}-{\mathbb E}({\bf x})) ({\bf x}-{\mathbb E}({\bf x}))^\top) $
 is shift-invariant, i.e.,
${\rm cov} ({\bf x}) {\bf S} = {\bf S} {\rm cov} ({\bf x})$.
 This, together with  \eqref{GFT.def} and  \eqref{shiftinvariantfilter.Fourier},
 implies that
\begin{equation}\label{stationary.Fourier}
{\rm cov}({\bf x})= {\bf U} {\bf P}_{\bf x} {\bf U}^\top, \ \ {\rm  equivalently,} \ \
{\rm cov}(\widehat {\bf x})=  {\bf P}_{\bf x}
\end{equation}
for some diagonal matrix ${\bf P}_{\bf x}$ with the diagonal entries considered as the power spectral density of the stationary signal ${\bf x}$
\cite{Girault2015, Perraudin2017, Jian2022, Zheng2025}.
In our approach to shift‑invariant channel estimation proposed in this paper, we select
{\it nonstationary} source signals whose covariance matrix exhibits a
non‑diagonal structure in the graph Fourier domain; see Assumption \ref{mainassumption}.

\end{enumerate}

\section{Channel estimation and blind reconstruction }
\label{sec:G-Channel}

Let $V_N=\{1, \ldots, N\}$. 
In this paper, we make the following assumptions about the  unknown  channel ${\bf H}$, channel noises  ${\bf n}_m$, and  source signals ${\bf x}_m, 1\le m\le M$.   

\begin{assumption}\label{mainassumption}
\begin{itemize}
\item[{(i)}] The unknown channel  ${\bf H}$ is real and  shift-invariant   with spectral
 representation \eqref{shiftinvariantfilter.Fourier}.

\item[{(ii)}]  
The channel noises ${\bf n}_m, 1\le m\le M$, are i.i.d.
random variables normally distributed with mean $0$ and unknown variance $\sigma^2$, i.e.,  ${\bf n}_m\sim {\mathcal N}(\mathbf{0}, \sigma^2 {\bf I})$.

\item[{(iii)}] The source signals ${\bf x}_m$, $1\le m\le M$, are realizations of an
i.i.d.\ random vector ${\bf x}$ with zero mean and 
 covariance  matrix ${\rm cov}(\widehat{\bf x}) := [C_{\widehat{\bf x}}(n,n')]_{n,n'\in V_N}$ 
of its GFT  available to
the observer. 

\item[{(iv)}] The  source graph ${\mathcal S}=(V_N, R)$ 
is connected, 
 where  the edge set $R$ contains all distinct pairs $(n,n')\in V_N\times V_N$ with
$C_{\widehat {\bf x}}(n,n')\ne 0$, i.e., 
 \begin{equation}\label{sourcegraph.def}
R=\left\{(n,n'):\ C_{\widehat{\mathbf x}}(n,n')\neq 0 \right\}.
\end{equation}

\item[{(v)}]  The channel noises ${\bf n}_m, 1\le m\le M$, and the source signals ${\bf x}_{m'}, 1\le m'\le M$, are independent. 

\end{itemize}

\end{assumption}

From (iii) and (iv) in Assumption \ref{mainassumption},
the source signals are  nonstationary  and lack ``stationary" frequency components by \eqref{stationary.Fourier}, and
 the covariance matrix
${\rm cov}(\widehat{\bf x})$ of their GFTs is available to the observer
whenever its covariance matrix  ${\rm cov}({\bf x})$ is, since ${\rm cov}(\widehat {\bf x})={\bf U}^\top {\rm cov}({\bf x}) {\bf U}$.

\medskip

For the noisy observation ${\bf y}$ in 
\eqref{xy.time}, we obtain from \eqref{xy.fourier} and Assumption
\ref{mainassumption} that
\begin{equation}\label{inputoutput.fourier}
\mathbb{E}\,\widehat{\bf y} = \mathbf{0}
\quad\text{and}\quad
{\rm cov}(\widehat{\bf y})
= {\pmb \Gamma}\,{\rm cov}(\widehat{\bf x})\,{\pmb \Gamma}
  + \sigma^2 {\bf I},
\end{equation}
where ${\rm cov}(\widehat{\bf y})
:= [C_{\widehat{\bf y}}(n,n')]_{n,n'\in V_N}$ is the covariance matrix of
the GFT of the observation ${\bf y}$, and
${\pmb \Gamma} = {\rm diag}[\gamma(1),\ldots,\gamma(N)]$ is the frequency
response of the unknown shift‑invariant channel ${\bf H}$ defined in
\eqref{shiftinvariantfilter.Fourier}.

Since ${\bf y}_m, 1\le m\le M$, in \eqref{equ:model} are independent realizations of the noisy observation ${\bf y}$,  
the empirical mean  ${\mathbb E}_M\widehat {\bf y}:=[{\mathbb E}_M\widehat{y}(n)]_{n\in V_N}$
and  empirical covariance matrix 
${\rm cov}_M(\widehat {\bf y}):=[C_{\widehat {\bf y}, M}(n,n')]_{n, n'\in V_N}$ of their GFTs
provide high-probability approximations to the true 
mean ${\mathbb E}\widehat {\bf y} $ and the covariance matrix
${\rm cov}({\widehat {\bf y}})$ respectively, 
where $\widehat {\bf y}_m=[\widehat { y}_m(n)]_{n\in V_N}, 1\le m\le M$,
\begin{equation}\label{EMy.def} {\mathbb E}_M\widehat{y}(n)= M^{-1} \sum_{m=1}^M \widehat{y}_m(n), \ \ n\in V_N,\end{equation}
 and
\begin{equation}\label{coym.def}
C_{\widehat{\bf y}, M}(n,n')= \frac{1}{M} \sum_{m=1}^M \big(\widehat{y}_m(n)-{\mathbb E}_M \widehat{y}(n)\big) \,
(\widehat { y}_m(n')- {\mathbb E}_M\widehat{y}(n')\big) , \ \ n, n'\in V_N.  
\end{equation}
In particular,  under the assumption that the source signal has
finite kurtosis, there exists a positive constant $C_0$ such that
for any sufficiently small $\epsilon>0$,
\begin{equation}
\label{covarianceapproximation}
|{\mathbb E}_M\widehat{y}(n)|\le \epsilon\ \ {\rm and} \ \
|C_{\widehat {\bf y}, M}(n,n')- C_{\widehat {\bf y}}(n,n')|\le \epsilon
\end{equation}
hold for all $1\le n, n'\in V_N$ with probability at least $1-C_0M^{-1} \epsilon^{-2}$; see Proposition \ref{approx.prop} in the Appendix.

Based on \eqref{inputoutput.fourier}, which characterizes the covariance relationship between the source signals and the noisy observations in the graph Fourier domain, together with the approximation result \eqref{covarianceapproximation} for the mean and covariance of the noisy observations, we adopt a phase retrieval framework to estimate the magnitudes of the frequency responses of the shift-invariant channel ${\bf H}$ in Section~\ref{channelestiamtion.section}. Subsequently, in Section~\ref{channelestiamtion2.section}, we assign signs to the estimated frequency responses, thereby providing an estimate of the unknown shift-invariant channel ${\bf H}$. 
The approximation  errors  for our shift-invariant channel estimation are stated in Proposition \ref{frequencynormestimate.prop} and
Theorem~\ref{population.uniqueness.thm}. 
In Section~\ref{recovery.subsection}, we propose a blind deconvolution scheme for source signal recovery and show that the discrepancy between the resulting observation data and the  noisy observations is of the order of the noise level. Moreover, assuming that the shift-invariant channel preserves all frequency components, we prove that, with high probability, the signals recovered by our proposed blind deconvolution method are close to those reconstructed by applying the Wiener filter at every frequency; see Theorem~\ref{tildex.thm} and \eqref{blinddeconvolutionsingalestimate}.

\subsection{Magnitudes of frequency
responses of the unknown shift-invariant channel}
\label{channelestiamtion.section}

Our estimation of the magnitudes of the frequency responses
$\gamma(n)$, $n\in V_N$, starts from an entry‑wise reformulation of the
covariance relationship \eqref{inputoutput.fourier} between the source
signal ${\bf x}$ and the noisy observation ${\bf y}$ in the graph Fourier
domain:
\begin{equation}\label{Dij.eqn}
C_{\widehat {\bf y}}(n,n')=\left\{\begin{array}{ll}
 |\gamma(n)|^2 C_{\widehat {\bf x}}(n, n)+\sigma^2  &  \ {\rm if} \ n'=n\\
 \gamma(n)\gamma(n') C_{\widehat {\bf x}}(n,n') & \ {\rm if} \ n'\ne n,
 \end{array}\right.
\end{equation}
where $n, n'\in V_N$. 
Let  $R$ be the edge set of the source graph ${\mathcal S}$ given in \eqref{sourcegraph.def}. For all distinct $n, n'\in V_N$ with $(n, n')\in R$, 
we obtain from \eqref{Dij.eqn} that
\begin{equation}\label {mun.meq0}
\left\{\begin{array}{l}
 |\gamma(n)|^2  C_{\widehat {\bf x}}(n, n) - |\gamma(n')|^2  C_{\widehat {\bf x}}(n', n')
 = \alpha(n, n')\\
 |\gamma(n)| \, |\gamma(n')|= |\beta(n, n')|,
\end{array}\right.
\end{equation}
where
$$\alpha(n, n')= C_{\widehat {\bf y}}(n, n)- C_{\widehat {\bf y}}(n', n') \ \ {\rm and} \ \ 
\beta(n, n')= C_{\widehat {\bf y}}(n, n')/C_{\widehat {\bf x}}(n, n').$$
Solving the quadratic equation \eqref{mun.meq0} for $|\gamma(n)|$
and $|\gamma(n')|$ yields
\begin{equation}\label{mun.meq1}
|\gamma(n)|=\sqrt{\frac{  \sqrt{\mu(n, n')}+\alpha (n, n') }{2\, C_{\widehat {\bf x}}(n, n)}}
\ \ {\rm for \ all} \ \ n'\in V_N \ \ {\rm with}\ \ C_{\widehat {\bf x}}(n, n')\ne 0, 
\end{equation}
where
$$\mu(n, n')= 
4  C_{\widehat {\bf x}}(n, n)  C_{\widehat {\bf x}}(n', n') (\beta (n, n'))^2+(\alpha(n, n'))^2.
$$
By the connectedness of the source graph ${\mathcal S}$ in Assumption
\ref{mainassumption}, we observe that $|\gamma(n)|$ is uniquely determined
for all $n\in V_N$, provided that the covariance matrix
${\rm cov}({\widehat{\mathbf{y}}})$ of the GFT of the observation
$\mathbf{y}$ is {\em available} to the observer; however, in our setting,  we only have
access to its empirical covariance matrix with entries defined by 
\eqref{coym.def}.

We denote the estimated shift-invariant channel by  ${\bf H}_M$ with the spectral representation
\begin{equation}\label{estimate.Fourier} {\bf H}_M ={\bf U} \,{\pmb \Gamma}_M{\bf U}^\top
\end{equation}
for some diagonal matrix ${\pmb \Gamma}_M={\rm diag}[\gamma_M(1), \ldots, \gamma_M (N)]$.
Based on the observation \eqref{mun.meq1} and the approximation property \eqref{covarianceapproximation} of
the empirical covariance, we
propose the following  estimates of the magnitudes of the
frequency responses $\gamma(n)$, $n\in V_N$,
and the noise level $\sigma$:
\begin{equation}\label{method1.mag}
|\gamma_M(n)|=\Bigg(\frac{1}{\theta(n)}
 \sum_{(n, n')\in R}
 \frac{
\sqrt{\mu_M(n, n')}+\alpha_M (n, n') } {
2 C_{\widehat {\bf x}}(n, n)}\Bigg)^{1/2}, \ \ n\in V_N,
\end{equation}
and
\begin{eqnarray}\label{noise.estimate}
\sigma_M^2 & = &  \frac{1}{N} \sum_{n\in V_N}
\big|C_{\widehat {\bf y}, M}(n, n)- |\gamma_M(n)|^2 C_{\widehat {\bf x}}(n, n)\big|\nonumber\\
& = & \frac{1}{2N}
 \sum_{n\in V_N} (\theta(n))^{-1} \Bigg|
 \sum_{(n, n')\in R} C_{\widehat {\bf y}, M}(n, n)+ C_{\widehat {\bf y}, M }(n', n')- \sqrt{\mu_M(n, n')}\Bigg|,
\end{eqnarray}
where $\theta(n)=\sum_{(n, n')\in R} 1$ is the degree of vertex $n\in V_N$ in  the source graph $\mathcal{S}$, and
\begin{equation}\left\{\begin{array}{l}  \alpha_M(n, n')= C_{\widehat {\bf y}, M}(n, n)- C_{\widehat {\bf y}, M }(n', n')\\
\beta_M(n, n')= C_{\widehat {\bf y}, M}(n, n')/C_{\widehat {\bf x}}(n, n')\\
 \mu_M(n, n')=(\alpha_M(n, n'))^2+ 
4  C_{\widehat {\bf x}}(n, n)  C_{\widehat {\bf x}}(n', n') (\beta_M (n, n'))^2.\end{array}\right. \end{equation}

The following proposition establishes large-deviation bounds for the estimation errors of the squared magnitudes of the frequency responses and the noise variance.

\begin{prop}\label{frequencynormestimate.prop}
Let $\sigma_M^2$ and  $|\gamma_M(n)|, n\in V_N$, be as in 
\eqref{noise.estimate} and \eqref{method1.mag}, 
$C_0$  be  the constant in \eqref{covarianceapproximation},
and 
let $\epsilon>0$ be chosen so that 
\begin{equation} \label{noise.epsilonreq} 16 D_0\epsilon\le \sigma^2,
\end{equation}
where we set
\begin{equation}
    \label{ConstantD.def}
    D_0=  \max_{(n, n')\in R} \frac{ \sqrt{ C_{\widehat {\bf x}}(n, n)  C_{\widehat {\bf x}}(n', n')}}{ |C_{\widehat {\bf x}}(n, n')| }\ge 1. \end{equation}
    Then 
\begin{equation}\label{frequencymagntitudeerror}
\big||\gamma_M(n)|^2-|\gamma(n)|^2\big|
 \le 3D_0 (C_{\widehat {\bf x}}(n, n))^{-1}
\epsilon, \ \ n\in V_N,
\end{equation}
and 
\begin{equation}\label{noiseestimate}
|\sigma_M^2-\sigma^2|\le (3D_0+1)\epsilon \le 4D_0 \epsilon 
\end{equation}
hold with probability at least $1-C_0 M^{-1}\epsilon^{-2}$.
\end{prop}

\begin{proof}
Set 
\begin{eqnarray*} g(t) & = & \Big( (C_{\widehat {\bf x}}(n, n'))^2 (t\alpha(n, n')+(1-t)\alpha_M(n, n'))^2\\
& & + \, 4 \, C_{\widehat {\bf x}}(n, n)  C_{\widehat {\bf x}}(n', n')  (tC_{\widehat {\bf y}}(n, n')+(1-t)C_{\widehat {\bf y}, M}(n, n'))^2\Big)^{1/2}.\end{eqnarray*} 
By \eqref{covarianceapproximation}, \eqref{mun.meq1} 
and \eqref{method1.mag}, we conclude that  the following error estimate about
the magnitudes of frequency responses of the unknown shift-invariant channel ${\bf H}$:
\begin{eqnarray*} 
\big||\gamma_M(n)|^2-|\gamma(n)|^2\big|
& \le  & 
\frac{1}{\theta(n)}
 \sum_{(n, n')\in R}
 \frac{ |\sqrt{\mu_M(n, n')}-\sqrt{\mu(n, n')}|+|\alpha_M (n, n')-\alpha(n, n')| } {
2 C_{\widehat {\bf x}}(n, n)}\nonumber\\
 & \le & \
 \max_{(n, n')\in R}
 \frac{ |C_{\widehat {\bf x}}(n, n')|^{-1} |g(1)-g(0)|
+|\alpha_M (n, n')-\alpha(n, n')| } {
2 C_{\widehat {\bf x}}(n, n)}\nonumber\\
& \le &  (D_0+2)  (C_{\widehat {\bf x}}(n, n))^{-1}
\epsilon \le 3D_0 (C_{\widehat {\bf x}}(n, n))^{-1}
\epsilon, \ \ n\in V_N,
\end{eqnarray*}
holds with probability at least $1-C_0 M^{-1}\epsilon^{-2}$,
where  the  third inequality follows from 
$|\alpha_M(n, n')-\alpha(n, n')|\le 2\epsilon$
and
\begin{eqnarray*} |g(1)-g(0)|
& \le &  
\sup_{a, b>0} \frac{ 2 a (C_{\widehat {\bf x}}(n, n'))^2 +
4 C_{\widehat {\bf x}}(n, n)  C_{\widehat {\bf x}}(n', n') b }
{ ((C_{\widehat {\bf x}}(n, n'))^2 a^2+
4 C_{\widehat {\bf x}}(n, n)  C_{\widehat {\bf x}}(n', n') b^2)^{1/2}} \epsilon \\
& \le & 2 \Big(  |C_{\widehat {\bf x}}(n, n')| +  \sqrt{ C_{\widehat {\bf x}}(n, n)  C_{\widehat {\bf x}}(n', n')}\Big) \epsilon 
\end{eqnarray*}
by \eqref{covarianceapproximation}. This proves \eqref{frequencymagntitudeerror}.

Following a similar argument, we obtain \eqref{noiseestimate} from \eqref{covarianceapproximation} and \eqref{frequencymagntitudeerror}.
\end{proof}

\subsection{Signs of frequency
responses of the unknown shift-invariant channel}
\label{channelestiamtion2.section}
By \eqref{Dij.eqn}, we have
\begin{equation}\label{Dijnondiagonal2.eqn} {\rm sgn}(\gamma(n))\,{\rm sgn}(\gamma(n')) = {\rm sgn}\!\left(\frac{C_{\widehat{\bf y}}(n,n')}{C_{\widehat{\bf x}}(n,n')}\right) = {\rm sgn}(\beta(n,n')), \quad \forall (n,n') \in R. \end{equation} Here  the sign function is defined by  ${\rm sgn}(x)= x/|x|$ for $x\ne 0$ and ${\rm sgn}(0)=0$.
In this subsection, we derive a procedure for assigning the signs of $\gamma_M(n)$, $n \in V_N$. The proposed procedure is built upon two key ingredients: the covariance approximation property in \eqref{covarianceapproximation} and the identities in \eqref{Dijnondiagonal2.eqn}, together with the notions of the perfect observation graph $\mathcal{D}$ and its empirical counterpart ${\mathcal D}_M$.

Define the  relative correlation matrix of the GFTs of  the noisy observation ${\bf y}$ by
 \begin{equation}\label{eq:rho_pop}
 \rho(n,n'):=\frac{|C_{\widehat {\bf y}}(n,n')|}{\sqrt{
 C_{\widehat {\bf y}}(n,n) C_{\widehat {\bf y}}(n',n') }}=
\frac{|\gamma(n)| |\gamma(n')| |C_{\widehat {\bf x}}(n ,n')|}
 {\sqrt{(|\gamma(n)|^2 C_{\widehat {\bf x}}(n, n)+\sigma^2)
 (|\gamma(n')|^2 C_{\widehat {\bf x}}(n', n')+\sigma^2)}}
 \end{equation}
for all $n, n'\in V_N$,
and set
\begin{equation} \label{observationgraph.def}
w(n,n')=\rho(n, n') \ {\rm if}\  n'\ne n\ {\rm and} \ w(n, n')=0 \ {\rm if} \  n'=n. 
\end{equation}
Let ${\mathcal D}=(V_N, B, W)$ be 
 the {\em perfect observation graph} 
with $V_N$ as the vertex set and 
 $W=[w(n, n')]_{n, n'\in V_N}$ as the weight matrix, and 
$C_k$, $1\leq k\leq K$, be the connected components of the perfect observation graph $\mathcal D$. 
It then follows from \eqref{Dijnondiagonal2.eqn} that the signs of the frequency responses of the unknown shift-invariant channel are uniquely determined on each connected component $C_k$, $1 \leq k \leq K$, up to a global sign ambiguity on every connected component.

Clearly, the perfect observation graph ${\mathcal D}$ (with edge weights ignored)
is a subgraph of the source graph ${\mathcal S}=(V_N,R)$, and the two graphs
coincide  if the shift-invariant channel ${\bf H}$ does not
annihilate any mode of variation, or equivalently,
\begin{equation} \label{uniqueness.thm.eq1} \gamma(n)\neq0,\ \   n\in V_N.\end{equation}
 Under this assumption on the channel $\mathbf{H}$ and the connectedness of the
source graph ${\mathcal S}$ in Assumption~\ref{mainassumption}, the signs of
the frequency responses of the unknown shift‑invariant channel $\mathbf{H}$
are fully determined, up to a global sign ambiguity, provided that the covariance
matrix of the GFT of the observation $\mathbf{y}$ is available to the observer.
In our setting, only an empirical estimate of the covariance matrix for the GFT
of the observation $\mathbf{y}$ is accessible. This motivates the following
approach to assign the signs of $\gamma_M(n), n\in V_N$.

Take  a positive constant  $\delta$ appropriately chosen, and define 
the empirical relative correlation matrix of the GFTs of observation data
${\bf y}_m, 1\le m\le M$, by
\begin{equation} \rho_M(n, n')= \frac{|C_{\widehat {\bf y}, M}(n,n')|}{\sqrt{
 C_{\widehat {\bf y}, M}(n,n) C_{\widehat {\bf y}, M}(n',n') }} \ \ {\rm for \ distinct }\ \ n, n'\in V_n.
 \end{equation}
 Let ${\mathcal  D}_M:=(V_N, B_M, W_M)$ 
be the {\em observation graph}
with the vertex set $V_N$ and weighted matrix $W_M=[w_M(n, n')]_{n, n'\in V_N}$ defined by
\begin{equation} \label{observation.edge.weight}
w_M(n, n')=\left\{ \begin{array}{ll} \rho_M(n, n')  & {\rm if}   \ \rho_M(n, n') \ge \delta \ {\rm and}\ (n, n')\in R\\
0  & {\rm otherwise}, 
\end{array}\right.
\end{equation}
and 
$C_{k, M}$, $1\leq k\leq K_M$, be the connected components of the  observation graph ${\mathcal D}_M$.
By \eqref{covarianceapproximation}, \eqref{frequencymagntitudeerror} and \eqref{noiseestimate}, we see that for any $\epsilon$ satisfying \eqref{noise.epsilonreq},    the following estimate
 \begin{equation} 
 \label{rhoM.approx} |\rho_M(n, n')-\rho(n, n')|\le 4 \epsilon \sigma^{-2}
 \end{equation}
 hold for all $n, n'\in V_N$
 with probability at least $1-C_0 M^{-1} \epsilon^{-2}$.
As a consequence, with $\delta>0$ appropriate chosen,
the conclusion
\begin{equation} \label{sameconnectedcomponent}
w_M(n, n')\ne 0 \ {\rm if \ and \ only\ if} \ \rho(n, n')\ne 0\end{equation}
hold for all $(n, n')\in R$ 
with probability at least
$1-64C_0 \sigma^{-4} M^{-1} \delta^{-2}$. In particular, we  may select $\delta>0$ so that
\begin{eqnarray}\label{delta.selection}
4 \epsilon \sigma^{-2} < \  \delta & \le &  
\frac{1}{3} \Big(\min_{\gamma(n)\ne 0} |\gamma(n)|^2\Big)
\inf_{(n, n')\in R}
\frac{|C_{\widehat {\bf x}}(n, n')|}
{\sqrt{ (\|{\bf H}\| C_{\widehat {\bf x}}(n, n))^2+\sigma^2}
\sqrt{ (\|{\bf H}\| C_{\widehat {\bf x}}(n', n'))^2+\sigma^2}
}\nonumber\\
& \le &  
\frac{1}{3} \inf\limits_{\substack{\ (n, n')\in R, \\
\ \rho(n, n')\ne 0}} \rho(n, n')\le \frac{1}{3D_0},
\end{eqnarray}
where
$\|{\bf H}\|=\max_{n\in V_N} |\gamma(n)|$, and $D_0$ is the constant in \eqref{ConstantD.def}.
Therefore, for sufficiently large $M$,  we conclude that, with high probability, 
the weight matrices of the perfect observation graph ${\mathcal D}$ and the  observation graph ${\mathcal D}_M$ have {\em small} differences  and they have exactly the {\bf same} connected components.

  By \eqref{Dijnondiagonal2.eqn}, \eqref{rhoM.approx} 
  and 
  the approximation property \eqref{covarianceapproximation} of
the empirical covariance,
the signs of the frequency responses $\gamma_{ M}(n),  n\in  V_N$, of the unknown shift-invariant channel ${\bf H}$ should be determined by using the 
following relation:
\begin{equation}\label{sign.req}
 \mathrm{sign} ( \gamma_{M}(n) ) \  \mathrm{sign} ( \gamma_{M}(n') )= \mathrm{sign}\big(\beta_M(n, n')\big) \  {\rm for \ all\  edges}\   (n,n')\in   B_{M}.
 \end{equation}
To mitigate sign inconsistencies, we assign the sign of $\gamma_M(n)$ for each $n \in V_N$ such that the condition \eqref{sign.req} is satisfied over a designated subset of edges, rather than across the entire  edges of the observation graph ${\mathcal D}_M$. 
To select such a subset of edges, we  decompose
the  observation graph  ${\mathcal D}_M$ into connected components
 $C_{k, M}, 1\le k\le K_M$,
 select a vertex $v_k$ from 
 the vertex set $W_{k, M}, 1\le k\le K_M$,
 of each connected component  $C_{k, M}$
 as the anchor vertex, and 
 choose  a traversal  $P_{k, M}$ that visits all vertices
 in $W_{ k, M}$ such that there is a unique path between every pair of vertices in $W_{k, M}$. 
 For the robustness of our approach, a common selection of the anchor vertex in the  connected component  $C_{k, M}$
  is
 the vertex with the maximum weighted degree, 
\[
v_k\in\operatorname*{arg\,max}_{n\in W_{k, M}}
\sum_{n'\ne n} w_M(n, n'),
\]
and the  traversal  $P_{k, M}$
in the connected component  $C_{k, M}$
is chosen so that it has maximal weights on the traversal, 
\[
P_{k, M}=\operatorname*{arg\,max}_{{\rm traversal} \ P\  {\rm in}\  C_{k, M}}
\sum_{(n, n')\in P} w_M(n, n').
\]
Then we assign an arbitrary sign for  the frequency response $\gamma_M(v_k)$
on  the anchor vertex of the  connected component $C_{k, M}$, i.e.,
 \begin{equation} \label{signrecovery.eq1}
 \mathrm{sign} ( \gamma_{M}(v_k) ) \in \{-1, 1\} \ \ {\rm for \ all} \ \ 1\le k\le K_M,
 \end{equation}
 and then, determine the 
 sign for  the frequency response $\gamma_M(n)$
on non-anchor vertices   of the  connected component $C_{k, M}$ by
 \begin{equation} \label{signrecovery.eq2}
 \mathrm{sign} ( \gamma_{M}(n) ) \  \mathrm{sign} ( \gamma_{M}(n') )= \mathrm{sign}\big(\beta_M(n, n')\big) \  {\rm for \ all}\   (n,n')\in   P_{k, M};
 \end{equation}
 cf. \eqref{sign.req}.
 The above sign assignment on each connected component
 of the approximate observation graph is well-defined since 
  $\beta_M(n, n')\ne 0$ on any edge $(n, n')$ in the traversal $P_{k, M}$ and  there is a unique path to connect any given non-anchor vertex $n$  to the anchor vertex $v_k\in W_{k, M}$.
  With the magnitude and sign of the frequency response of the
shift-invariant filter ${\bf H}_M$ given in \eqref{method1.mag},
\eqref{signrecovery.eq1}, and \eqref{signrecovery.eq2}, we have  an
estimation of the  shift-invariant channel ${\bf H}$. 
The above procedure is summarized in the Covariance-based
Shift-Invariant Channel Estimation Algorithm (CSICE); see
Algorithm~\ref{alg1:channel-est}.

\begin{algorithm}[t]
	\begin{algorithmic}
		\STATE INPUT:   Covariance matrix
       ${\rm cov}(\widehat {\bf x})$ of the nonstationary source signal in the graph Fourier domain, 
       observations ${\bf y}_m, 1\le m\le M$, and empirical covariance 
    $C_{\widehat {\bf y}, M}$ 
    in \eqref{coym.def},  
the thresholding constant  $\delta$,
and  the  observation graph ${\mathcal D}_M=(V_N, B_M, W_M)$
in \eqref{observation.edge.weight}.

\STATE ALGORITHM: 
\begin{enumerate}[label=\textnormal{(\roman*)}]

\item Use \eqref{method1.mag} to define the magnitude 
$|\gamma_{M}(n)|$, $n\in V_N$, of the frequency response 
 the estimated shift-invariant channel 
${\bf H}_M$.

\item Find all connected components $C_{k, M}, 1\le k\le K_M$ of the observation  graph ${\mathcal D}_M$ and denote the set of vertices in each connected component by
$W_{k, M}, 1\le k\le K_M$.

\item Find an anchor vertex $v_k$ in each connected component $C_{k, M}$, and assign arbitrary sign $\epsilon_k\in 
\{-1, 1\}$ to the frequency response  $\gamma_M(v_k)$ 
of the estimated shift-invariant channel ${\bf H}_M$.

\item For each  connected component $C_{k, M}$, find a traversal 
$P_{k, M}$ in $C_{k, M}$ to connect all vertices in $W_{k, M}$.

\item For $1\le k\le K$,  use \eqref {signrecovery.eq2} on the traversal $P_{k, M}$ to determine the signs of $\gamma_M(n)$
at non-anchor vertices $n\in W_{k, M}$.

\end{enumerate}

		\STATE OUTPUT:  The  estimated shift-invariant channel  ${\bf H}_{ M}={\bf U} \mathbf{\Gamma}_M {\bf U}^\top$ with $\mathbf{\Gamma}_M={\rm diag}(\gamma_M(1), \ldots, \gamma_M(N))$.
\end{algorithmic}
\caption{Covariance-based Shift-Invariant Channel Estimation Algorithm (CSICE)}
	\label{alg1:channel-est}
\end{algorithm}

\smallskip 

\begin{rem} {\rm 
Let 
\[
V_{N,M}
   = \Bigl\{\, n\in V_N :\  \rho_M(n,n') > \delta 
      \ \text{for some } n'\neq n \Bigr\}
\]
be the set of all vertices of the observation graph 
${\mathcal D}_M$ that have at least one incident edge.
Then, any vertex $n\notin V_{N,M}$ is isolated either because the magnitude of 
$\gamma_M(n)$ is extremely small, or because the original source
signal ${\bf x}$ contains a ``stationary'' component at frequency
$n$ in the sense that the channel has a tiny frequency response
at all vertices connected to $n$ in the source graph.  
For vertices $n\notin V_{N,M}$, the  corresponding connected component has only one element $\{n\}$. Consequently,  an
arbitrary sign is assigned to the corresponding frequency response
of the shift-invariant channel ${\bf H}$
in our approach.
}
\end{rem}

By \eqref{frequencymagntitudeerror}, \eqref{sameconnectedcomponent}, \eqref{signrecovery.eq1} and \eqref{signrecovery.eq2}, 
 the spectral responses  of the estimated channel ${\bf H}_M$ closely approximate the true spectral responses  of the shift-invariant channel ${\bf H}$, up to a sign ambiguity in each connected component of the observation graph, as the sample size $M$ is  large, i.e., 
there exists $\epsilon_k\in \{-1, 1\}$ such that
\begin{equation}\label{response.estimate}
    \gamma_M(n)\approx \epsilon_k  \gamma(n), \ \ n\in W_{k, M}.
\end{equation}
Therefore, more like a general phase retrieval problem \cite{Fienup82, jaganathan2016, Fannjiang2020}, 
the shift-invariant channel ${\bf H}_M$, with frequency response
$\gamma_M(n)$, $n\in V_N$, obtained in Algorithm~\ref{alg1:channel-est},
provides a good approximation to either ${\bf H}$ or $-{\bf H}$ with high
probability when the sample size $M$ is large, provided that the
shift-invariant channel ${\bf H}$ does not annihilate any mode of
variation, and the thresholding constant $\delta$ used to define the
observation graph is chosen appropriately.

\begin{thm}
\label{population.uniqueness.thm}
Consider the blind deconvolution problem
\eqref{equ:model} satisfying Assumption~\ref{mainassumption}.
Assume that all frequency responses of the shift-invariant channel ${\bf H}$ do not vanish, i.e,  \eqref{uniqueness.thm.eq1} holds. 
Select the thresholding constant 
$\delta\le  \inf_{(n,n')\in R} \rho(n, n')/2$, define the observation graph ${\mathcal D}_M$ as in \eqref{delta.selection}, and let 
${\bf H}_M$ be the shift-invariant channel with its
frequency estimate  $\gamma_M(n), n\in V_N$, given by 
\eqref{method1.mag},
\eqref{signrecovery.eq1} and \eqref{signrecovery.eq2}.
Then,
there exists $s\in \{-1, 1\}$ such that 
the following  inequalities
\begin{equation}  \label{uniqueness.thm.eq2}
 |\gamma_M(n)-s \gamma(n)|
\le \frac{ 3D_0} 
 { C_{\widehat {\bf x}}(n, n) |\gamma(n)|}
\epsilon, \ \ n\in V_N
\end{equation}
hold  with probability at least $1-C_0M^{-1}\epsilon^{-2}$, where
$C_0$  and $D_0$ are the constants in \eqref{covarianceapproximation} 
and \eqref{ConstantD.def} respectively, 
and
\begin{equation}\label{epsilon.req2} \epsilon\le \frac{1}{6D_0}\min\Big( D_0 \delta \sigma^2, \ \min_{n\in V_N} |\gamma(n)|^2 C_{\widehat {\bf x}}(n, n)\Big). \end{equation}
\end{thm}

\begin{proof}
With   the selection of the thresholding constant $\delta$
in \eqref{delta.selection}, 
the observation graph ${\mathcal D}_M$ (without taking the weight into account) is the same as the perfect observation graph ${\mathcal  D}$ with probability at least $1-64 C_0M^{-1}\delta^{-2}\ge 1-C_0M^{-1} \epsilon^{-2}$ by \eqref{sameconnectedcomponent}.
This, together with the observation that the perfect observation graph
(without taking the weight into account) is the same as the source graph by 
the assumption on the shift-invariant channel ${\bf H}$, implies that the observation graph ${\mathcal D}_M$ (without taking the weight into consideration) is the same as the source graph ${\mathcal S}$, and hence, it is connected by 
 Assumption \ref{mainassumption}.
Therefore, there is only one connected component in the observation graph ${\mathcal D}_M$ and 
the proposed shift-invariant channel ${\bf H}_M$
is unique up to a global sign ambiguity. Therefore, the 
desired conclusion \eqref{uniqueness.thm.eq2} follows from  \eqref{frequencymagntitudeerror} and the above uniqueness.
\end{proof}

\subsection{Blind deconvolution of nonstationary graph signals}
\label{recovery.subsection}

By  Proposition  \ref{frequencynormestimate.prop} and 
Theorem~\ref{population.uniqueness.thm}, the noise level 
$\sigma_M$ in \eqref{noise.estimate} and the shift-invariant channel
${\bf H}_M$ with frequency estimates $\gamma_M(n)$, $n\in V_N$, obtained
from \eqref{method1.mag}, \eqref{signrecovery.eq1} and
\eqref{signrecovery.eq2} provide reasonable estimates of the noise level
$\sigma$ and the unknown shift-invariant channel ${\bf H}$ (up to a global ign ambiguity), with
high probability. 
Therefore, we may reformulate our model 
\eqref{equ:model} approximately as
\begin{equation}\label{approximate:model}
 {\bf y}_m\approx  {\bf H}_M  {\bf x}_{m}+ {\bf n}_{m, M}, \ 1\le m\le M, 
\end{equation}
where  ${\bf n}_m$ are i.i.d.\ Gaussian noise samples following the normal
distribution ${\mathcal N}(0,\sigma_M^2 {\bf I})$.
Applying GFT to both sides of the above approximation model yields
\begin{equation}\label{approximate:model}
\widehat{y}_m(n) \approx  \gamma_M(n) \widehat{x}_{m}(n)+ \widehat{n}_{m, M} (n), \ n\in V_N, 
\end{equation}
for $1\le m\le M$.
Based on the above observation, we propose the following blind deconvolution  $\widetilde {\mathbf{x}}_{m}$, $1 \le m \le M$, whose GFTs ${\widehat {\widetilde {\bf x}}}_m= [ {\widehat{ \widetilde{x}}}_{m}(n)]_{n\in V_N}$ are defined by
\begin{equation}\label{blindreconstruction.eq}
{\widehat{ \widetilde{x}}}_{m}(n)
= 
\frac{\gamma_M(n)\, C_{\widehat{\mathbf{x}}}(n,n)}
     { C_{\widehat {\bf y}, M}(n,n)}
\widehat{y}_{m}(n)\ \ 
{\rm for }\  n \in V_N \ {\rm and}\ \ 1\le m\le M,
\end{equation}
where the Wiener filter is applied independently at each frequency,
disregarding any correlation between the source signals across different
frequencies.

Let $\widetilde{\mathbf{y}}_{m}$, $1 \le m \le M$, be the reconstructed
observation defined by
\begin{equation}\label{tildeym.def}
\widetilde{\mathbf{y}}_{m} = \mathbf{H}_{M}\,\widetilde{\mathbf{x}}_{m},
\ \ 1 \le m \le M,
\end{equation}
and define the differences between the reconstructed observations and the noisy
observations by
\begin{equation}\label{observationdifference.def}
\mathbf{y}_{m}^{*} = \widetilde{\mathbf{y}}_{m} - \mathbf{y}_{m},
\ \ 1 \le m \le M.
\end{equation}
In the following theorem, we establish concentration estimates for the empirical mean and the empirical covariance matrix of the differences $\mathbf{y}_{m}^{*}$, $1 \le m \le M$.

 \begin{thm}
\label{tildex.thm}
Consider the blind deconvolution problem
\eqref{equ:model} satisfying Assumption~\ref{mainassumption}.
 Let
 $\widetilde{\mathbf{x}}_m, \widetilde{\mathbf{y}}_m$ and ${\bf y}_m^*, 1\le m\le M$, be the reconstructed source signals in 
\eqref{blindreconstruction.eq},  
 the reconstructed observation data in \eqref{tildeym.def}, and
 the  differences between the reconstruction observation data and the noisy observations in \eqref{observationdifference.def}, respectively.
 Denote the empirical mean and empirical covariance matrix
of the GFTs of ${\bf y}^*_m, 1\le m\le M$, by
 ${{\mathbb E}_M \widehat{ {\bf y}^*}} = [ {\mathbb E}_M\widehat{ {\mathbf y}^*}(n)]_{n\in V_N}$
 and the
empirical covariance matrix ${\rm cov}_M\big(\widehat{{\mathbf y}^*}\big)=[C_{{\widehat{ {\bf y}^*}}, M}(n, n')]_{n, n'\in V_N}$, respectively.
Take $\epsilon>0$  so that  \eqref{noise.epsilonreq} holds.
 Then 
the following estimates hold,
\begin{equation}\label{tildex.thm.eq1}
|{\mathbb E}_M\widehat{y^*}(n)|
 \le \epsilon, \  \ n\in V_N, 
\end{equation}
and 
\begin{equation}\label{tildex.prop.eq2}
|C_{{\widehat{ {\bf  y}^*}}, M}(n, n')|
 \le    2\epsilon+
    \frac{ 2 |C_{\widehat{\bf y}}(n, n'))|}
  {  |C_{\widehat{\bf y}}(n, n)) |C_{\widehat{\bf y}}(n', n'))|
} \sigma^4, \ \ n, n'\in V_N, 
    \end{equation}
    with probability at least $1-C_0 M^{-1} \epsilon^{-2}$, 
where  
$C_0$ is the positive constant in \eqref{covarianceapproximation}, and
$C_{\widehat {\bf y}}(n, n')$, $n, n'\in V_N$, is given in \eqref {Dij.eqn}.
 \end{thm}

\begin{proof} 
Let ${\bf y}_m, 1\le m\le M$, be the noisy observations in 
the blind deconvolution problem
\eqref{equ:model} and 
${\mathbb E}_M \widehat{\bf y} = [ {\mathbb E}_M\widehat{y}(n)]_{n\in V_N}$
be the empirical mean
 \eqref{EMy.def} of their GFTs. 
Clearly, we obtain from \eqref{blindreconstruction.eq} and \eqref{tildeym.def}
that 
\begin{equation} \label{tildex.thm.pf.eq1}
{\mathbb E}_M\widehat{y^*}(n)= - \nu_M(n)
 {\mathbb E}_M {\widehat{y}}(n),
    \end{equation}
where $\nu_M(n)= \left(C_{\widehat{\bf y}, M}(n, n) -   \gamma_M(n)^{2}\, C_{\widehat{\mathbf{x}}}(n,n)\right)/
C_{\widehat {\bf y}, M}(n), n\in V_N$.
Then the desired estimate  \eqref{tildex.thm.eq1} follows from   
\eqref{covarianceapproximation}, \eqref{frequencymagntitudeerror}, \eqref{noise.epsilonreq} and 
\eqref{tildex.thm.pf.eq1}.

By \eqref{blindreconstruction.eq} and \eqref{tildeym.def}, we have
\begin{equation} \label{tildex.thm.pf.eq2}
C_{\widehat{{\bf y}^*},M}(n,n')
=
\nu_M(n)  \nu_M(n')
      C_{\widehat{{\mathbf{y}}},M}(n,n'), \ \
n, n'\in V_N.  
\end{equation}
From \eqref{frequencymagntitudeerror} and \eqref{noise.epsilonreq}, it follows that 
\begin{equation} \label{tildex.thm.pf.eq3}
\Big|  |\gamma_M(n)|^2C_{\widehat{\bf x}}(n, n)- |\gamma(n)|^2C_{\widehat{\bf x}}(n, n)\Big|\le \frac{3\sigma^2}{16}, \ n\in V_N, \end{equation}
  hold with probability at least $1-C_0 M \epsilon^{-2}$.
Combining 
\eqref{covarianceapproximation}, \eqref{Dij.eqn}, \eqref{tildex.thm.pf.eq2} and
\eqref{tildex.thm.pf.eq3},
we conclude that the following inequalities hold  with probability at least $1-C_0 M \epsilon^{-2}$,
\begin{eqnarray} \label{tildex.thm.pf.eq4}   
|C_{\widehat{\bf y}^*,M}(n,n')|
  & \le &  \epsilon+
  \frac{ (\sigma^2+ 4D_0\epsilon)^2 |C_{\widehat{\bf y}}(n, n')|}
  {
  ( C_{\widehat{\bf y}}(n, n)-\epsilon) ( C_{\widehat{\bf y}}(n', n')-\epsilon)}\nonumber\\
 &   \le  & 
  2\epsilon+
    \frac{ 2 |C_{\widehat{\bf y}}(n, n'))|}
  {  |C_{\widehat{\bf y}}(n, n)) |C_{\widehat{\bf y}}(n', n'))|
} \sigma^4 , \ \ n. n'\in V_N.
 \end{eqnarray}
This proves \eqref{tildex.prop.eq2} and completes the proof.
\end{proof}

By Theorem \ref{tildex.thm}, we see that, with high probability, the empirical
mean of the GFTs of the differences $\mathbf{y}_m^{*}$, $1 \le m \le M$, is
close to zero. Moreover, the covariance $C_{{\widehat{ {\bf  y}^*}}, M}(n, n')$ between the  distinct $n$-th and $n'$-th frequency components
of the difference $\mathbf{y}_m^{*}$ for $(n,n') \notin R$ is close to zero, while for
$(n,n') \in R$ or $n'=n$, their covariance is bounded by $3\sigma^{2}$ with high
probability, since
\[
2\epsilon +
\frac{2\, |C_{\widehat{\mathbf{y}}}(n,n')|}
     {|C_{\widehat{\mathbf{y}}}(n,n)|\, |C_{\widehat{\mathbf{y}}}(n',n')|}
     \sigma^{4}
\le \frac{\sigma^{2}}{16} + 2\sigma^{2}
\le 3\sigma^{2}.
\]
This demonstrates the reasonableness of the proposed blind deconvolution estimate $\widetilde{\mathbf{x}}_{m}$, $1 \le m \le M$, in \eqref{blindreconstruction.eq}, as the resulting data-fitting error cannot be distinguished from the intrinsic noise level~$\sigma$.

\medskip

By \eqref{xy.fourier}, we may  apply 
the Wiener filter  independently at each frequency, disregarding any correlation between the source signals across different frequencies, to reconstruct 
source signals ${\bf x}_{m, W}, 1\le m\le M$, whose GFTs are defined by 
\begin{equation}\label{wienerreconstruction.eq}
{\widehat{ {x}}_{m, W}}(n)
= 
\frac{\gamma(n)\, C_{\widehat{\mathbf{x}}}(n,n)}
     {(\gamma(n))^{2}\, C_{\widehat{\mathbf{x}}}(n,n)
      + \sigma^{2}}
\,\widehat{y}_{m}(n) \  \ {\rm for }
\ \ n \in V_N \ {\rm and}\ \ 1\le m\le M.
\end{equation}
Define the differences between the  proposed blind deconvolution and (negative)  Wiener  recovery by 
\begin{equation}\label{signaldifference.def}
{\bf x}_m^{*, s} = \widetilde {\bf x}_m- s{\bf x}_{m, W}, \ \  1\le m\le M\end{equation}
for $s\in \{-1, 1\}$, and denote its GFT by 
$\widehat{{\bf x}_m^{*, s}}=[\widehat {{\bf x}_m^{*, s}}(n)]_{n\in V_N} $.

 If the shift-invariant channel ${\bf H}$ does not annihilate any mode of variation, i.e., \eqref{uniqueness.thm.eq1} holds, then we obtain from \eqref{frequencymagntitudeerror},  \eqref{noiseestimate}, \eqref{noise.epsilonreq} and  
 \eqref{uniqueness.thm.eq2} that 
\begin{eqnarray}\label{blinddeconvolutionsingalestimate}
|\widehat {{\bf x}_m^{*, s}}(n)| & = & \Bigg |
\frac{\gamma_M(n)\, C_{\widehat{\mathbf{x}}}(n,n)}
     {C_{\widehat {\bf y}, M}(n, n)}- \frac{s\gamma(n)\, C_{\widehat{\mathbf{x}}}(n,n)}
{C_{\widehat {\bf y}}(n, n)}     \Bigg| |\widehat {y}_m(n)|\nonumber\\
      & \le & \Big(\frac{3D_0 \, \epsilon}
      {|\gamma(n)|   C_{\widehat {\bf y}}(n, n)
  }+\frac{ 32 \, \epsilon \, |\gamma(n)| \, C_{\widehat{\bf x}}(n, n)}
            { 15 ( C_{\widehat {\bf y}}(n, n))^2 }\Big)  |\widehat {y}_m(n)|\nonumber\\
      & \le & \frac{5 D_0 \, \epsilon}
      {|\gamma(n)|     ((\gamma(n))^{2}\, C_{\widehat{\mathbf{x}}}(n,n)
      + \sigma^{2}) }|\widehat {y}_m(n)|, \ 1\le m\le M,      
\end{eqnarray}
hold with probability at least $1-C_0M^{-1} \epsilon^{-2}$, 
 where $s\in \{-1, 1\}$ is the sign in Theorem \ref{population.uniqueness.thm}.
 This result indicates that, with high probability, the proposed blind deconvolution approximates the Wiener deconvolution at each frequency, up to a global sign ambiguity.

\section{ Numerical Demonstrations}
\label{demonstration.section}

In this section, we demonstrate the performance of the proposed covariance-based channel estimation  for shift-invariant channels and the blind deconvolution framework introduced in the previous section. We present two numerical demonstrations: one based on synthetic data on cycle graphs and another using temperature dataset collected in the Brest region of France.  The details of these experiments are provided in Sections \ref{sec:circulant} and \ref{temperature.section} respectively.

\subsection{Blind deconvolution on cycle graphs}
\label{sec:circulant}

Take $N\ge 3$, and let ${\mathcal C}_N$ be the unweighted cycle graph with vertex set $V_N$ and edge set
$\{(n, n\pm 1 \ {\rm mod}\  N): \ n\in V_N\}$. 
In the simulations of this section, we select the  Laplacian on the cycle graph ${\mathcal C}_N$
as the graph shift ${\bf S}$, take an
 orthogonal matrix ${\bf U}$ so that 
${\bf S}=\mathbf{U}\,\mathrm{diag}(\lambda(1),\ldots,\lambda(N))\mathbf{U}^\top$ for its eigenvalues $0=\lambda(1)<\lambda(2)\le \ldots \le \lambda(N)\le 2$, and define the GFT of a graph signal ${\bf x}$ on ${\mathcal C}_N$ by 
$\widehat {\bf x}={\bf U}^\top  {\bf x}$.

In our  simulations, we  assume that the  unknown shift-invariant channel ${\bf H}$
in our blind deconvolution problem \eqref{equ:model} is of the form
\begin{equation}\label{cyclegraph.eq1} {\bf H}={\bf U} {\pmb \Gamma} {\bf U}^\top \end{equation}
for some diagonal matrix ${\pmb \Gamma}={\rm diag}[\gamma(1), \ldots, \gamma(N)]$ with 
the magnitudes $|\gamma(n)|, n\in V_N$, being independently drawn from a uniform distribution on $[0.5,2]$ and the signs being chosen independently at random with equal probability, 
and the nonstationary source signal ${\bf x}$ having covariance matrix given by
\begin{equation} \label{cyclegraph.eq2} {\rm cov}({\bf x})={\bf U} (\alpha {\bf I}+\beta {\bf S}) {\bf U}^\top,
\end{equation}
where $\alpha\ge 0$ and $\beta>0$.
Under the above assumption on the source signals, we observe that 
the source graph ${\mathcal S}$
is the same as the cycle graph ${\mathcal C}_N$ (and hence, the requirement (iv) in Assumption \ref{mainassumption} is met), and the traversal  used to determine the sign
of the frequency response is the unique path in the connected component, except that in the scenario that the observation graph is the same as the source graph.

\begin{figure}[t] 
\includegraphics[width=0.43\linewidth]{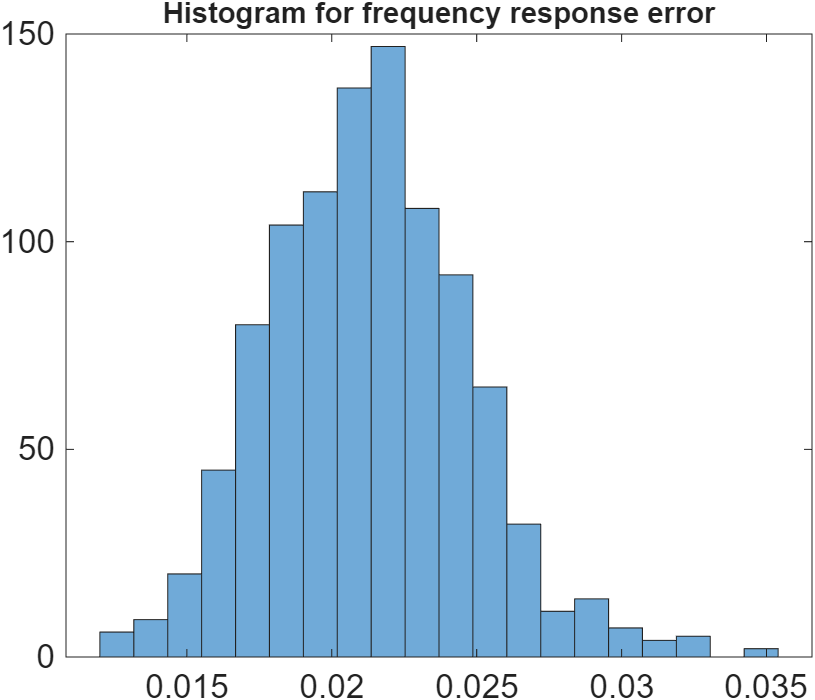}
\includegraphics[width=0.43\linewidth]{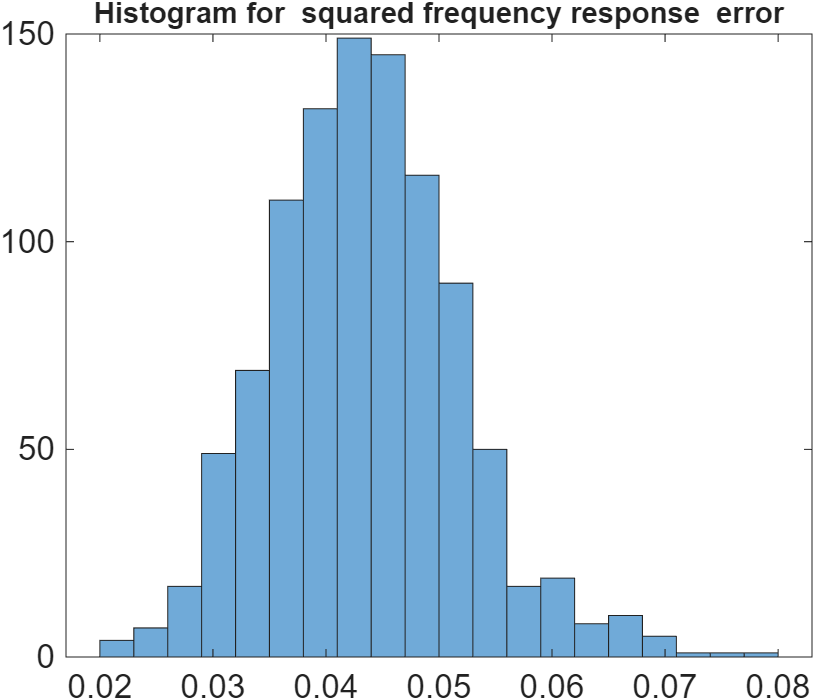}
\caption{Histograms of the  average  error \(E_1\) between the true and estimated frequency responses of the shift-invariant channel (left), and the average error \(E_2\) between the squared true and estimated frequency responses (right), where $N=30, M=5000, \alpha=3, \beta=1$ and $\sigma=0.2$. The histograms are obtained from $1000$ independent trials.  The errors $E_1$ and $E_2$ have mean values $0.0214$ and $0.0440$, respectively.
}
\label{magnitudeErrorHisogram.fig}
\end{figure}

By \eqref{cyclegraph.eq1} and \eqref{cyclegraph.eq2}, we observe that 
\begin{equation} \label{cyclegraph.eq3} {\rm cov}
(\widehat {\bf x})=\alpha {\bf I}+\beta {\bf S} \ \ {\rm and}\ \  {\rm cov}(\widehat {\bf y})= {\pmb \Gamma}
(\alpha {\bf I}+\beta {\bf S}) {\pmb \Gamma}+\sigma^2 {\bf I}. 
\end{equation}
Therefore the constant  $D_0$  in \eqref{ConstantD.def}
is given by    $D_0= (\alpha+2\beta)/\beta$ and
the thresholding constant $\delta$ in \eqref{delta.selection} can be chosen to be
$\delta = \beta a^2 (3(\alpha+2\beta)a^2+3\sigma^2)^{-1}$,
since 
\begin{eqnarray*} \min\limits_{\substack{(n, n')\in R,\\ \rho(n, n')\ne 0}} \rho(n, n') & = & 
\inf\limits_{\substack{\gamma(n), \gamma(n')\ne 0,\\ n'=n\pm 1 \ {\rm mod}\ N }} \frac{\beta |\gamma(n)| |\gamma(n')|}
{\sqrt{ (\alpha+2\beta)|\gamma(n)|^2+\sigma^2} \sqrt{ (\alpha+2\beta)|\gamma(n')|^2+\sigma^2} }\nonumber\\
& \ge &  
\frac{\beta a^2 }{ (\alpha+2\beta)a^2+\sigma^2}, 
\end{eqnarray*}
provided that  $|\gamma(n)|\ge a$ for all $n\in V_N$.
Thus for our proposed channel estimate ${\bf H}_M$ with frequency estimate $\gamma_M(n)$, $n\in V_N$, given by \eqref{method1.mag} \eqref{method1.mag},
\eqref{signrecovery.eq1} and \eqref{signrecovery.eq2},
and  for sufficiently small $\epsilon>0$, 
 it follows from  Theorem \ref{population.uniqueness.thm} that 
the inequalities
\begin{equation}\label{cyclegraph.eq3} |\gamma(n)| \, |\gamma_M(n)-s\gamma(n)| \le 3\beta^{-1}\epsilon \end{equation}
and
\begin{equation}\label{cyclegraph.eq4}|(\gamma_M(n))^2- (\gamma(n))^2| \le 3\beta^{-1}\epsilon,  \ \ n\in V_N \end{equation}
 hold with probability at least $1-C_0M^{-1}\epsilon^{-2}$ for some $s\in \{-1, 1\}$.
 Shown in Figure~\ref{magnitudeErrorHisogram.fig} are histograms of the average errors: 
\begin{equation}\label{Error1.eq} E_1:=N^{-1} \min_{s\in \{-1, 1\}} \sum_{n\in V_N} |\gamma_M(n)-s\gamma(n)|
\ \  {\rm and}\ \ 
E_2:=N^{-1} 
\sum_{n\in V_N} |(\gamma_M(n))^2- (\gamma(n))^2|,\end{equation} 
 where \(E_1\) measures the discrepancy between the true frequency response \(\gamma(n)\), \(n\in V_N\), of the shift-invariant channel \({\bf H}\) and the  frequency response \(\gamma_M(n)\) of the shift-invariant channel estimate \({\bf H}_M\), up to a global sign ambiguity, while \(E_2\) measures the discrepancy between their squared frequency responses.

\begin{figure}[t] 
\includegraphics[width=0.43\linewidth]{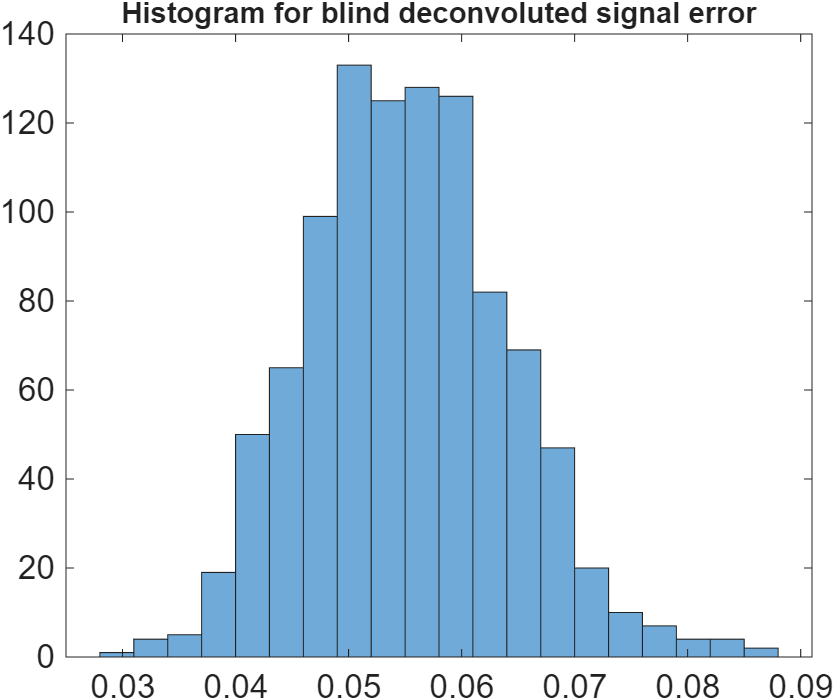}
\includegraphics[width=0.43\linewidth]{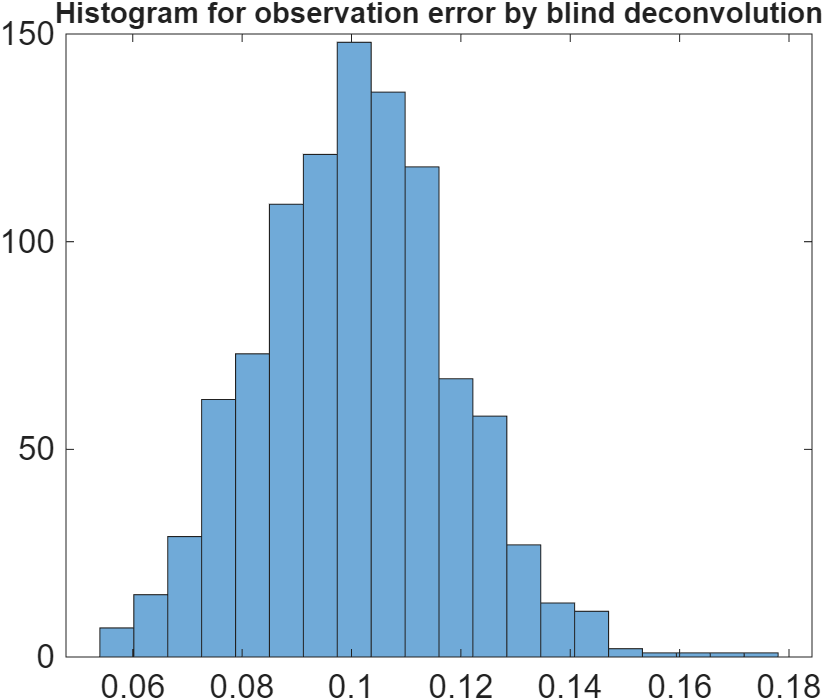}
\caption{Histograms of the blind deconvolution error $E_3$ (left) and observation error $E_4$ (right) with mean values $0.0549$ and $0.1014$ respectively, computed from $1000$ independent trials,  where  $N=50, M=5000, \alpha=3, \beta=1$ and $\sigma=0.2$.}
\label{Sourceerror.fig}
\end{figure}

Let $\widetilde {\mathbf{x}}_{m}$, $1 \le m \le M$,
be the  signals in  \eqref{blindreconstruction.eq}
reconstructed from our proposed  blind deconvolution  approach and 
${\bf x}_{m, W}, 1\le m\le M$, be as in   \eqref{wienerreconstruction.eq} reconstructed from applying  Wiener filtering in the Fourier domain at each frequency.
Let $\widetilde {\bf y}_m, 1\le m\le M$, be the reconstructed observation in \eqref{tildeym.def}.
Combining \eqref{blinddeconvolutionsingalestimate} with Theorem~\ref{tildex.thm}, we conclude that, with high probability, 
$\widetilde {\mathbf{x}}_{m}$ approximates either 
${\bf x}_{m, W}$ or $-{\bf x}_{m, W}, 1\le m\le M$, and 
the difference between $\widetilde {\bf y}_m$ and ${\bf y}_m, 1\le m\le M$ is at the noise level.
Shown on the left and right of  Figure~\ref{Sourceerror.fig} are histograms of the approximate errors
\[ E_3 = \inf_{s\in\{-1,1\}} \frac{1}{M} \sum_{m=1}^{M} \bigl\| \widetilde{\mathbf{x}}_m - s\mathbf{x}_{m,W} \bigr\|_2 = \inf_{s\in\{-1,1\}} \frac{1}{M} \sum_{m=1}^{M} \bigl\| \widehat{\widetilde{\mathbf{x}}}_m - s\widehat{\mathbf{x}}_{m,W} \bigr\|_2 \] 
between the blind-deconvolution estimate and the Wiener estimate of the source signals at each frequency after resolving the global sign ambiguity,
and 
\[ E_4 = \frac{1}{M} \sum_{m=1}^{M} \bigl\| \widetilde{\mathbf{y}}_m-\mathbf{y}_m \bigr\|_2 = \frac{1}{M} \sum_{m=1}^{M} \bigl\| \widehat{\widetilde{\mathbf{y}}}_m - \widehat{\mathbf{y}}_m \bigr\|_2 \] 
 between the true observation and the output obtained by filtering the deconvolved source signals through the proposed shift-invariant channel, 
over $1000$ independent trials.

\subsection{Blind deconvolution of temperature data}
\label{temperature.section}
In  this  subsection, we demonstrate the performance
of the blind deconvolution method
\eqref{blindreconstruction.eq} for the 
hourly temperature dataset
collected at 32 weather stations in the Brest region (France)  during the month of January 2014 \cite{Perraudin2017, 
Zheng2025}.
We represent the dataset of size $32\times 24\times 31$
by ${\bf x}_{d, t}^{\rm org} $,
representing the regional temperature at 
 hour $t$ of day  $d$ in January 2014, where $0\le t\le 23$ and $1\le d\le 31$.   
 In our simulations, we pre-process the temperature dataset by substracting the average hourly temperature recorded
 in 
 January 2014, 
 i.e.,
 $${\bf x}_{d, t}= {\bf x}_{d, t}^{\rm org} - \frac{1}{31}\sum_{d'=1}^{31} {\bf x}_{d', t}^{\rm org}  \ \ {\rm for}\   1\le d \le 31 \ {\rm and}\  0\le t\le 23.$$
The above centered temperature  dataset
${\bf x}_{d, t}$ 
has zero  
 mean, thereby allowing it to be interpreted as  a collection of  realizations of a zero-mean random process.

\begin{figure}[t] 
	\centering
    \includegraphics[width=0.48\linewidth]{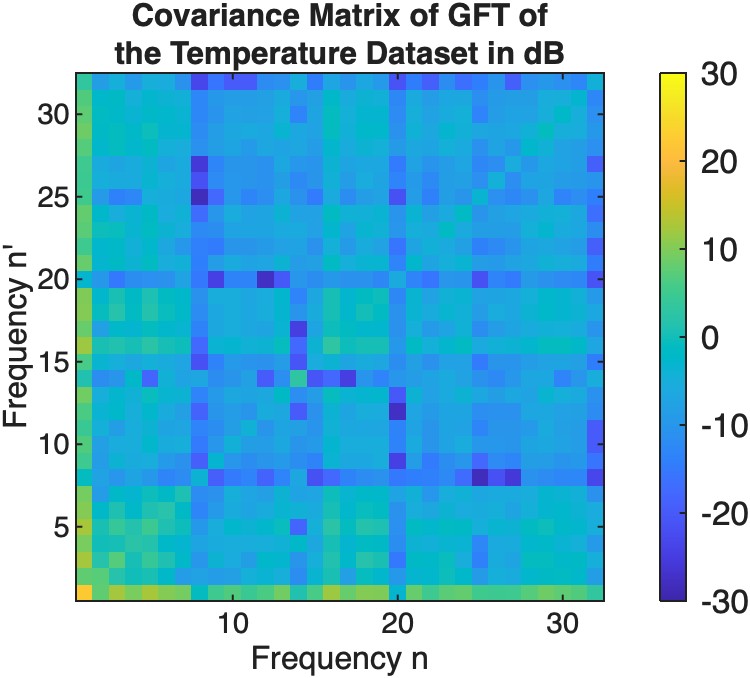} \ \
        \includegraphics[width=0.48\linewidth]{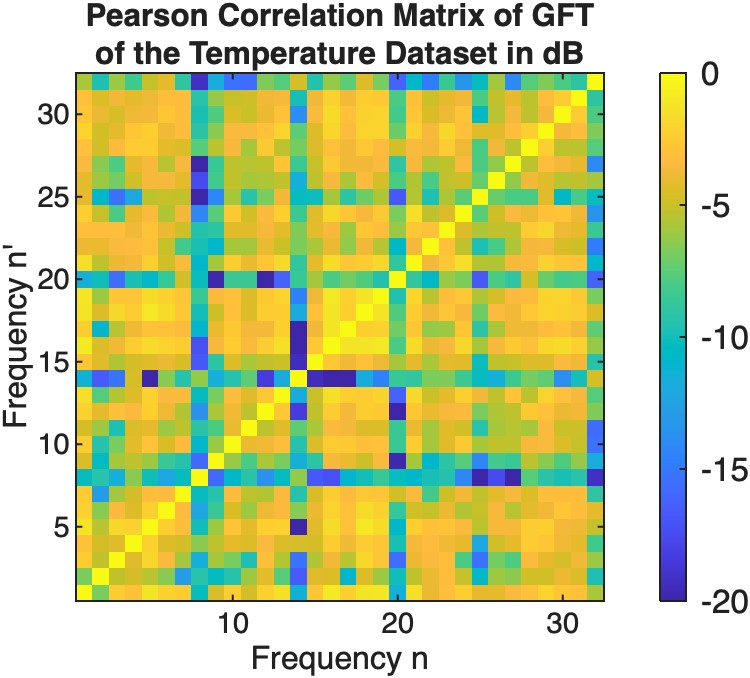}    

	\caption{
    The covariance matrix (left) and Pearson correlation matrix (right) of the GFTs of the centered temperature dataset ${\bf x}_{d,t}$, $1 \le d \le 31$, $0 \le t \le 23$, shown on the dB scale.
 }
	\label{temperatureCovariance.fig}
\end{figure}

 In our simulation, we use the same underlying graph ${\mathcal G}$ as in \cite[Figure 10]{Perraudin2017}, which
is essentially built from the coordinates of 32 weather stations by 
connecting all the neighbors in a given radius. In our simulations, we use the Laplacian on the graph ${\mathcal G}$
 as the graph shift ${\bf S}$ to define GFT; see \eqref{GFT.def}.  Set $V_{32}=\{1, 2, \ldots, 32\}$, and define the 
 covariance matrix  ${\bf C}_{\widehat {\bf x}}=[C_{\widehat{\mathbf{x}}}(n,n')]_{n, n'\in V}$ of GFTs of the centered temperature dataset  by 
 \begin{equation}\label{covariance.sim}
{C}_{\widehat{\mathbf{x}}}(n,n') =
\frac{1}{31\times 24} 
\sum_{d=1}^{31}\sum_{t=0}^{23} {\widehat x}_{d, t}(n) 
{\widehat x}_{d, t}(n'), \ \  n, n'\in V_{32}.
\end{equation}
The variance $C_{\widehat {\bf x}}(n, n), n\in V_{32}$,   
has value  $177.8017$
at zero frequency ($n=1$), 
and the rest exhibits a 
range from $5.1511$ 
to  $0.2584$.  
The covariance matrix $\max(10\log_{10} (\lvert {C}_{\widehat{\mathbf{x}}}(n,n') \rvert+10^{-5}), -30),  n,n'\in V_{32}$
is displayed on the decibel (dB) scale on the left plot of  Figure \ref{temperatureCovariance.fig}.  
It is observed that the centered temperature dataset exhibits properties characteristic of nonstationary graph signals, making it
suitable serving as  the source signals in our simulations; see Assumption~\ref{mainassumption}.

We select the source graph ${\mathcal S}=(V_{32}, R_T)$ of the centered temperature dataset
 based on its spectral
 Pearson correlation coefficients 
$$\rho_{\widehat{\bf x}}(n,n') = \frac{C_{\widehat{\mathbf{x}}}(n,n')}{\sqrt{C_{\widehat{\mathbf{x}}}(n,n)} \sqrt{C_{\widehat{\mathbf{x}}}(n', n')}},\ \  n, n'\in V_{32}.$$
Shown in the right plot of Figure 
\ref{temperatureCovariance.fig}
is 
the spectral Pearson correlation $\max(-20,$ $10 \log (|\rho_{\widehat{\bf x}}(n,n')|+10^{-5}))$ on the dB scale.  
In our simulations, we select the pairs $(n, n')\in V_{32}\times V_{32}$
with  $|\rho_{\widehat {\bf x}}(n,n')|\ge 0.01$ as edges in the source graph ${\mathcal S}$.  
Then, the edge set $R_T$ of the source graph ${\mathcal S}$ contains all undirected pairs of distinct vertices except for the following six pairs $(5, 14), (8, 25), (8, 27), (12,20 )$, $(14, 16), (14, 17)$. In other words, the source graph ${\mathcal S}$ differs from a complete graph on 32 vertices by only six missing undirected edges associated with negligible correlations. As a result, ${\mathcal S}$ is considerably denser than the cycle graph considered in the previous subsection, in which each vertex is connected solely to its two neighboring vertices.

  For an unknown shift-invariant channel ${\bf H}$ with frequency responses $\gamma(n), n\in V_{32}$, the noisy observations ${\bf y}_{d, t}$  have their  GFTs represented by
 $${\widehat y}_{d, t}(n)=\gamma(n) \, \widehat x_{d, t}(n)+ \sigma n_{d, t} (n), \ \ n\in V_{32}, 1\le d\le 31\ \ {\rm and}\ \ 0\le t\le 23,$$
 where   $\sigma>0$ is the noise level, and ${\bf n}_{d,t}(n)$ are i.i.d white noises with standard normal distribution. 
 
In our first simulation, we consider the scenario in which
the unknown shift-invariant channel preserves all frequency components, i.e., \eqref{uniqueness.thm.eq1} holds.
In particular, the shift-invariant channel ${\bf H}$
is assumed to have  frequency responses 
\begin{equation}\label{allpass.simulation1} \gamma(n)= \epsilon(n) \times (1+u(n)),\  n\in V_{32},\end{equation} where $u(n), n\in V_N$ are
independently drawn from a uniform distribution on $[-0.2, 0.2]$,
and  the signs $\epsilon(n)\in\{-1, 1\}, n\in V_{32}$, are  chosen
independently at random with equal probability.
 Following the procedure in Section \ref{channelestiamtion.section}, we construct the observation graph ${\mathcal D}$ with thresholding constant $\delta=0.001$. It is observed that, for $\sigma\le 1/2$, (i)  the observation graph ${\mathcal D}$  either coincides with the source graph ${\mathcal S}$, or differs from it by the deletion of only a small number of edges;
(ii) the observation graph ${\mathcal D}$ is connected; and
(iii) the vertices corresponding to components with nonzero frequencies have an edge connecting to the vertex $n=1$ corresponding to the component with zero frequency.  Then, we set $M=31\times 24$,  use 
 \eqref{method1.mag} to estimate the magnitude of 
 frequency response, $|\gamma_M(n)|, n\in V_{32}$,
 assign positive sign for $\gamma_M(1)$,  apply
 \eqref{signrecovery.eq2} for the  traversal with edges $(1, n), 1\ne n\in V_{32}$
 to determine the signs of $\gamma_M(n), 1\ne n\in V_{32}$,
 and obtain  frequency responses $\gamma_M(1), \ldots, \gamma_M(32)$
 of the shift-invariant channel estimate ${\bf H}_M$. 
 Therefore, the GFT of the recovered signals  $\widetilde {\bf x}_{d, t}$ using our method \eqref{blindreconstruction.eq} is given by 
 \begin{equation}\label{reconstructed.sim}
 {\widehat {\widetilde  x}}_{d, t}(n) = \frac{\gamma_M(n) C_{\widehat {\bf x}}(n, n)}{C_{\widehat{\bf y}}(n, n)}
 \widehat y_{d, t}(n) \ {\rm for} \ n\in V_{32},  1\le d\le 31\  {\rm and} \ 0\le t\le 23,
 \end{equation}
 where
 $$ 
 C_{\widehat{\bf y}}(n, n)= M^{-1} \sum_{d=1}^{31} \sum_{t=0}^{23}
  (\widehat y_{d, t}(n))^2- \Big(M^{-1} \sum_{d=1}^{31} \sum_{t=0}^{23}
  \widehat y_{d, t}(n)\Big)^2, \ \ n\in V_{32}. $$

\begin{figure}[t] 
	\centering

\begin{subfigure}{.48\linewidth}
\centering
\includegraphics[width=\linewidth]{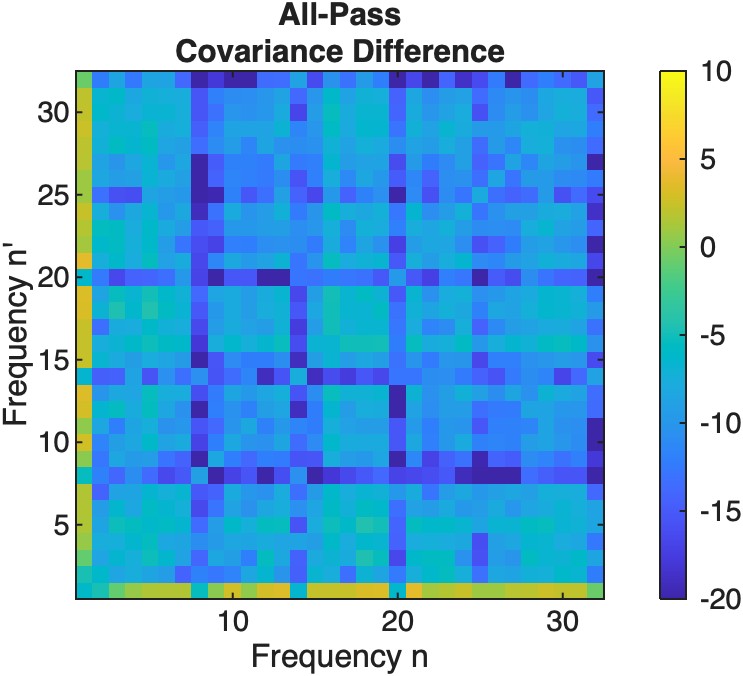}\vspace{-4pt}
\end{subfigure}\hfill
\begin{subfigure}{.48\linewidth}
\centering
\includegraphics[width=\linewidth]{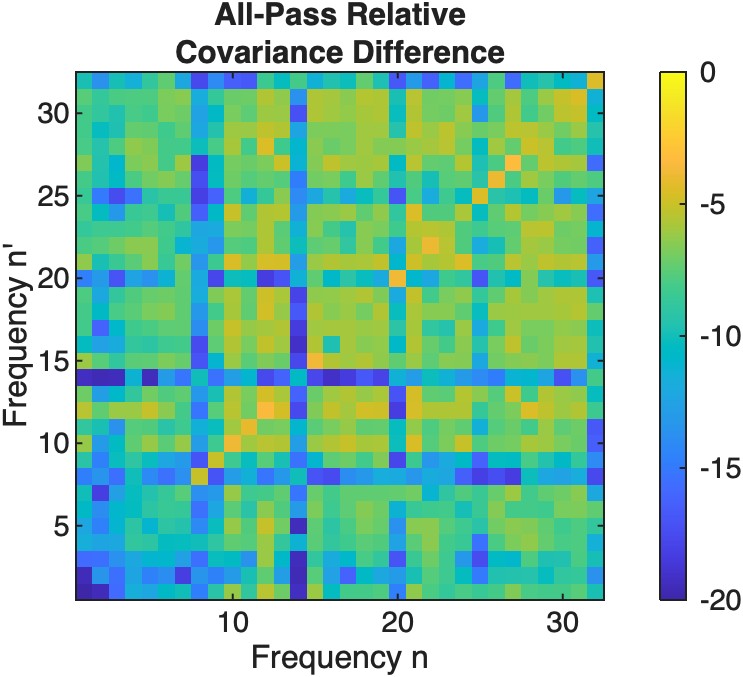}\vspace{-4pt}
\end{subfigure}\\[4pt]
\vspace{10pt}
\begin{subfigure}{.48\linewidth}
\centering
\includegraphics[width=\linewidth]{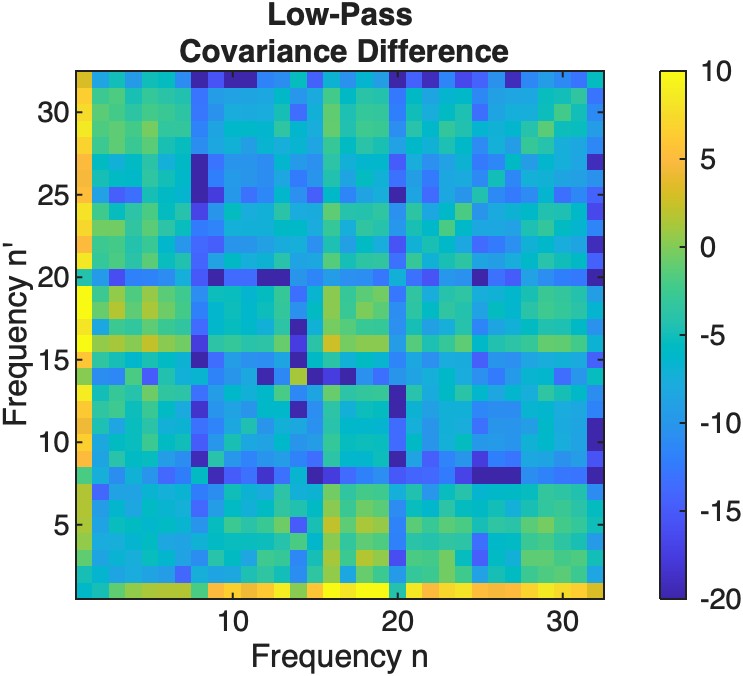}\vspace{-4pt}
\end{subfigure}\hfill
\begin{subfigure}{.48\linewidth}
\centering
\includegraphics[width=\linewidth]{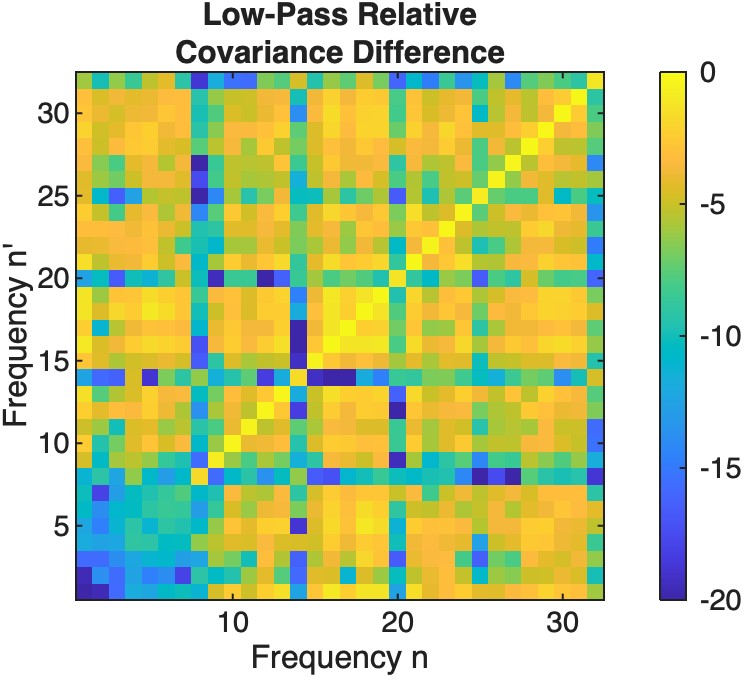}\vspace{-4pt}
\end{subfigure}\\[4pt]
\vspace{-4pt}
\caption{Covariance matrix difference (left) and relative covariance matrix difference (right), in dB scale, between the GFTs of the centered temperature dataset $\mathbf{x}_{d,t}$ and those of the blind-deconvolved dataset $\widetilde{\mathbf{x}}_{d,t}$, where $1 \le d \le 31$ and $0 \le t \le 23$, averaged over $1000$ trials. The noise level is set to $\sigma=1/2$, and the color ranges are fixed to $[-20,10]$ dB for the plots in the left column and $[-20,0]$ dB for the plots in the right column. The top row corresponds to the shift-invariant channel $\mathbf{H}$ given by the all-pass filter in \eqref{allpass.simulation1}, while the bottom row corresponds to the low-pass filter in \eqref{lowpass.simulation}.
 }
    \label{temperatureCovarianceDiff.fig}
\end{figure}

Denote the empirical covariance matrix of ${\widehat {\widetilde  {\bf x}}}_{d, t}$
in \eqref{reconstructed.sim} by
${\bf C}_{{\widehat {\widetilde {\bf x}}}, M}=[C_{{\widehat {\widetilde  {\bf x}}}, M}(n, n')]_{n, n'\in V_{32} } $, where
$$  {\mathbb E}_{{\widehat{\widetilde {\bf x}}}, M}(n)= 
 M^{-1} \sum_{d=1}^{31} \sum_{t=0}^{23}
 { \widehat {\widetilde x}}_{d, t}(n), \ n\in V_{32},
$$
and 
$$ C_{{\widehat{\widetilde {\bf x}}}, M}(n, n')= M^{-1} \sum_{d=1}^{31} \sum_{t=0}^{23}
 \big ({\widehat{\widetilde x}}_{d, t}(n)- {\mathbb E}_{{\widehat{\widetilde {\bf x}}}, M}(n)\big)  \big ({\widehat{\widetilde x}}_{d, t}(n')- {\mathbb E}_{{\widehat{\widetilde {\bf x}}}, M}(n')\big), \ \ n, n'\in V_{32}. $$
 By 
\eqref {blinddeconvolutionsingalestimate}, the above  empirical covariance matrix 
${\bf C}_{{\widehat {\widetilde {\bf x}}},M} $ should be a good approximation to the reference covariance matrix ${\bf C}_{\widehat {\bf x}} $ defined in \eqref{covariance.sim} for the centered temperature dataset.  Presented at the top row of  Figure \ref{temperatureCovarianceDiff.fig} is the empirical evaluation of this approximation over 1000 trials, where the average of the covariance difference
$$\min\Big(10, \max\big(-20, 10*\log_{10}(|C_{{\widehat {\widetilde  {\bf x}}}, M}(n, n')- C_{\widehat {\bf x}}(n, n')|+10^{-5})\big)\Big)$$
(left) and relative covariance difference
$$\max(10*\log_{10}(|C_{{\widehat {\widetilde  {\bf x}}}, M}(n, n')- C_{\widehat {\bf x}}(n, n')|/ \sqrt{C_{\widehat {\bf x}}(n, n) C_{\widehat {\bf x}}(n', n')}+10^{-5}), -20), \ n,n'\in V_{32},$$
(right) both in the dB scale and with noise level $\sigma=1/2$, are displayed. 

We observe that the covariance difference is significantly larger at the lowest frequency component ($n=0$) than at the remaining frequencies, whereas the relative covariance difference remains at a similar level across all other frequency components.
The possible reason is that the  temperature dataset  has a large  component at the first frequency, as verified that $C_{\widehat{\mathbf{x}}}(1,1) = 177.8$ 
 is considerably larger than $C_{\widehat{\mathbf{x}}}(n,n)\in [0.2584, 5.1511], 2\le n\le 32$.

In the second simulation, we consider that the shift-invariant channel is essentially a low-pass filter.
We observe that eigenvalues of the graph Laplacian (also used as the graph shift) have a visible jump at $n=7$. 
In the second simulation, we assume that the unknown shift-invariant channel ${\bf H}$ has the same first seven frequency-response coefficients, $\gamma(n)$, $1 \leq n \leq 7$, as those used in the first simulation, while all the remaining frequency-response coefficients are set to zero, i.e.,
\begin{equation}\label{lowpass.simulation}
    \gamma(n)= \left\{\begin{array}{ll} \epsilon(n)(1+u(n))  & {\rm if} \ 1\le n\le 7\\
    0 & {\rm otherwise},
    \end{array}\right.
\end{equation}
$u(n), n\in V_N$ are
independently drawn from a uniform distribution on $[-0.2, 0.2]$,
and  the signs $\epsilon(n)\in\{-1, 1\}, n\in V_{32}$, are  chosen
independently at random with equal probability.

Denote by ${\bf H}_M$ the estimate of the shift-invariant channel ${\bf H}$, 
with its frequency responses $\gamma_M(n)$, $n \in V_{32}$, obtained from \eqref{method1.mag}, \eqref{signrecovery.eq1}, and \eqref{signrecovery.eq2}. Following the first simulation, we have the noisy observation ${\bf y}_{d, t}$ and blind-deconvolved signals $\widetilde {\bf x}_{d, t}, 1\le d\le 31$ and $0\le t\le 23$, and  the empirical covariance matrix
${\bf C}_{{\widehat{\widetilde {\bf x}}}, M}$. Shown in the bottom
 row of  Figure \ref{temperatureCovarianceDiff.fig} is the empirical evaluation of this approximation over 1000 trials, where showing the average of the covariance difference
(left) and relative covariance difference
(right) both expressed in the dB scale and with  noise level $\sigma=1/2$. 
From the bottom right plot of Figure \ref{temperatureCovarianceDiff.fig}, we observe that the relative covariance difference is close to zero over the low frequency range of the shift-invariant channel, indicating that the covariance information is well preserved and remains close to that of the original signal under the proposed blind deconvolution scheme.  On the other hand, the high frequency components are not adequately represented by the low pass channel and consequently exhibit larger covariance errors. This behavior is also reflected in the diagonal entries of the covariance difference matrix, where the low frequency components have substantially smaller errors than the higher frequency components.

\begin{appendix}

\vskip-0.02in


\section{Empirical covariances of source signals and noisy observations}
\label{approxproof.appendix}

In this appendix, we establish an error estimate for the empirical
covariances of the source signals and the noisy observations.

\begin{prop}\label{approx.prop}
 Let the  channel ${\bf H}$,  the source signals ${\bf x}_m$ and the noises ${\bf n}_m, 1\le m\le M$, satisfy
Assumption \ref{mainassumption}, and 
take $\epsilon>0$.
If the GFT of the source random variable  ${\bf x}$ has 
finite
kurtosis, i.e., 
$C_4:=\max_{n\in V_N} {\mathbb E} \,  |\widehat {\bf x}(n)|^4<\infty$,  
then
\begin{equation} \label{approx.prop.eq2}
{\rm Pr}\left( \lvert C_{\widehat {\bf x}, M}(n, n')- C_{\widehat {\bf x}}(n, n') \rvert \ge \epsilon \right) \le  \frac{C_4}{M\epsilon^2}
\end{equation}
for any given $n, n'\in V_N$; 
\begin{equation} \label{approx.prop.eq3}
{\rm Pr}\big(\left| C_{\widehat {\bf y}, M}(n, n)- C_{\widehat {\bf y}}(n, n) \right| \ge \epsilon \big)  
\le 
\frac{C_4\|{\bf H}\|^4 + 6\sqrt{C_4} \sigma^2 \|{\bf H}\|^2 + 3\sigma^4}
{M\epsilon^2}
\end{equation}
for any given $n \in V_N$; 
and
\begin{equation} \label{approx.prop.eq4}
{\rm Pr}\big(|C_{\widehat {\bf y}, M}(n, n')- C_{\widehat {\bf y}}(n, n')|\ge \epsilon\big)  
\le \frac{C_4\|{\bf H}\|^4 + 2\sqrt{C_4} \sigma^2 \|{\bf H}\|^2 + \sigma^4}{M\epsilon^2}
\end{equation}
for any given $n, n'\in V_N$ with $n\ne n'$,
where $\sigma^2$ is the channel noise  variance, and 
$\|{\bf H}\|=\sup_{\|{\bf x}\|_2=1} \|{\bf H}{\bf x}\|_2$ is the operator norm of the channel ${\bf H}$.
    \end{prop}

\begin{proof}
For $n, n'\in V_N$, we have
\begin{equation} \label{approx.prop.pfeq1}
 {\mathbb E}\  \big|\widehat x(n) \widehat x(n')- {\mathbb E}\big(\widehat x(n) \widehat x(n')\big) \big|^2 \le     {\mathbb E}\  | \widehat x(n) \widehat x(n') |^2
 \le
\max_{1\le \widetilde n\le N} {\mathbb E}  \ |\widehat x(\widetilde n) |^4\le C_4.
\end{equation}
Combining  the classical law of large numbers
with the variance estimates in
\eqref{approx.prop.pfeq1} proves the desired estimate
\eqref{approx.prop.eq2}.

For $n\in V_N$, we obtain from   \eqref{xy.fourier} and Assumption \ref{mainassumption} that
\begin{eqnarray} \label{approx.prop.pfeq2}
& & \hskip-0.08in  {\mathbb E} \big|(\widehat y(n))^2- {\mathbb E}\big(\widehat y(n))^2)\big |^2
\le  
{\mathbb E} |\widehat y(n) |^4= {\mathbb E}
\big(\gamma(n)\widehat x(n)+\widehat n(n)\big)^4 \nonumber\\
&\hskip-0.08in = & \hskip-0.08in 
|\gamma(n)|^4 {\mathbb E} |\widehat x(n)|^4+
6 \sigma^2 |\gamma(n)|^2  {\mathbb E} |\widehat x(n)|^2 + 3\sigma^4
\le 
\|{\bf H}\|^4 C_4+
6 \sigma^2 \|{\bf H}\|^2 \sqrt{C_4}+ 3\sigma^4.
\end{eqnarray}
This, together with the classical law of large numbers, proves   
\eqref{approx.prop.eq3}.

For $n, n'\in V_N$ with $n\ne n'$, we have
\begin{eqnarray*} \label{approx.prop.pfeq3}
& & \hskip-0.08in  {\mathbb E} \big|\widehat y(n) \widehat y(n')- {\mathbb E}\big(\widehat y(n) \widehat y(n')\big)\big |^2\le  {\mathbb E} \big | ((\mu(n)\widehat x(n)+\widehat n(n)) (\mu(n')\widehat x(n')+\widehat n(n'))\big |^2 \nonumber\\
&\hskip-0.08in    =  &  \hskip-0.08in (\mu(n))^2 (\mu(n'))^2  {\mathbb E} |\widehat x(n)|^2 |\widehat x(n')|^2
+ \sigma^2 (\mu(n))^2 {\mathbb E} |\widehat x(n)|^2+ \sigma^2 (\mu(n'))^2 {\mathbb E} |\widehat x(n')|^2+\sigma^4\nonumber
\\
& \hskip-0.08in \le & \hskip-0.08in
\|{\bf H}\|^4 C_4+
2 \sigma^2 \|{\bf H}\|^2 \sqrt{C_4}+ \sigma^4.
\end{eqnarray*}
This, in conjunction with the  law of large numbers, establishes  \eqref{approx.prop.eq4}.
\end{proof}

\end{appendix}

{\bf Acknowledgement}:\  
The authors acknowledge the assistance of Microsoft Copilot to provide phrasing refinements and typesetting suggestions.

{\bf Data availability}:  All data supporting the findings of this study are available from the corresponding author upon reasonable request.

{\bf Competing interests}:  The authors have no competing interests to declare that are relevant to the content
of this article.

\end{document}